\documentclass[a4paper,fleqn]{cas-dc}

\usepackage[numbers,sort&compress]{natbib}
\usepackage{amsmath,amssymb}
\usepackage{booktabs}
\usepackage{xcolor}
\usepackage{array}
\newcolumntype{Y}[1]{>{\raggedright\arraybackslash}p{#1}}
\usepackage{graphicx}
\usepackage{tikz}
\usetikzlibrary{arrows.meta,positioning}
\usepackage{pgfplots}
\pgfplotsset{compat=1.16}
\usepgfplotslibrary{groupplots}
\definecolor{cbblue}{HTML}{0072B2}
\definecolor{cbverm}{HTML}{D55E00}
\definecolor{cbgreen}{HTML}{009E73}
\definecolor{cborange}{HTML}{E69F00}
\definecolor{cbpurple}{HTML}{CC79A7}

\begin{document}
\let\WriteBookmarks\relax
\def\floatpagepagefraction{1}
\def\textpagefraction{.001}

\shorttitle{Fixed pilots can overstate the gains of learned receivers}
\shortauthors{M. Tasin et.al}

\title [mode = title]{Fixed pilot sequences can overstate the gains of
learned MIMO-OFDM receivers under power amplifier distortion}

\author[1]{Mohammad Tasin}
\cormark[1]
\ead{tasin2006072@gmail.com}

\author[2]{Al-Imran}
\ead{alimran1703075@gmail.com}

\affiliation[1]{organization={Department of Electrical and Electronic Engineering,
                Bangladesh University of Engineering and Technology},
            city={Dhaka},
            postcode={1205},
            country={Bangladesh}}

\affiliation[2]{organization={Department of Electronic Engineering, Kookmin University},
            city={Seoul},
            postcode={02707},
            country={Republic of Korea}}

\cortext[1]{Corresponding author}

\begin{abstract}
Neural-network receivers are proposed to undo the distortion of a power
amplifier driven close to saturation, where it is most efficient. They estimate the channel from pilots, known symbols sent in
every frame. The pilots pass through the same amplifier as the data, so
the channel estimate carries a small error that depends on the pilot
sequence. If one sequence is always sent, as by default in the
Sionna simulator, the error never changes, and a network can learn it
instead of undoing the amplifier. We tested this in
simulated two- and four-antenna orthogonal frequency-division multiplexing
links; results are for binary pilots at the highest signal-to-noise ratio
tested. A network with separate weights for each subcarrier lowered the bit
error rate by 14 to 22\% with its training pilot but raised it by 2 to
18\% with every other pilot tested, and longer training widened this gap.
A convolutional neural receiver following an open-source tutorial gained
37 percentage points more on its training pilot than on unseen pilots
after short training, and 9 points more after nine times longer training,
much of which also appeared with a linear amplifier. The gap shrank with amplifier back-off but persisted where the
in-band distortion meets the 5G limit. On a channel with short delay
spread, smoothing the channel estimate across subcarriers without
training lowered the error rate on all 107 pilot sequences tested, by
30\% on average.
Learned receivers should
therefore be tested on unseen pilots and against such simple baselines.
\end{abstract}


\maketitle

\section{Introduction}\label{sec:intro}

The power amplifier (PA) is one of the largest consumers of energy in a
radio transmitter, and it is most efficient when it is driven close to
saturation \citep{raab2002,lopezperez2022}. Close to saturation, however,
it compresses the strongest parts of the signal and shifts their phase
\citep{saleh1981,rapp1991}. Orthogonal frequency-division multiplexing
(OFDM), the waveform of 4G, 5G and Wi-Fi, suffers in particular: the sum
of many subcarriers produces occasional high peaks, and the distortion of
each peak spreads over all subcarriers \citep{dardari2000,han2005}. Each
classical remedy has a price. Backing the amplifier off from saturation
wastes power, peak-reduction methods change the transmitted signal
\citep{han2005}, digital predistortion needs a model of the amplifier and
a feedback path \citep{morgan2006}, and cancelling the clipping noise at
the receiver needs knowledge of the nonlinearity \citep{chen2003}.
Receivers built from neural networks are an alternative
\citep{oshea2017,ye2018,hoydis2021,shlezinger2023}. Convolutional
receivers detect strongly distorted OFDM signals
\citep{honkala2021deeprx,pihlajasalo2021hybridpimrc,pihlajasalo2022spawc,pihlajasalo2023},
training over several amplifier models makes them less sensitive to the
exact amplifier \citep{nielsen2023}, model-aided networks have been
combined with multiple-input multiple-output (MIMO) detection
\citep{xu2021}, and recent designs report large gains
\citep{ahmadian2025,kumar2026}.

Such gains are measured in simulation, and the way they are measured can
decide the result. Before it can detect
data, a receiver must estimate the channel, that is, how the signal on
each subcarrier was attenuated and rotated on its way from the
transmitter. It does this with pilots, symbols that the receiver knows in
advance. The simplest estimate, the least-squares (LS) estimate, divides
each received pilot symbol by the one that was sent. The pilots, however, pass through the same amplifier as
the data, so they arrive distorted \citep{ligata2011,xu2021}, and the
estimate is slightly wrong on every subcarrier. The error depends on the
exact pilot sequence. If a simulation sends the same sequence in every
frame, the error repeats in every frame, and a network trained in that
simulation can learn to cancel it. Tested with the same sequence, the
network then appears to have learned to undo the amplifier, while it has
in fact learned a property of one sequence, much as a student who
memorises the answers to one exam paper would fail another paper on the
same subject. Machine learning calls this shortcut learning
\citep{geirhos2020,degrave2021,lapuschkin2019}, and the overlap between
training and test conditions is a form of leakage \citep{kapoor2023}. It
matters because many systems change their pilots. In 5G New Radio (NR),
the demodulation reference signal depends on the slot number, the OFDM
symbol and a scrambling identity \citep{3gpp38211}, so a gain that exists
only for one pilot sequence would not appear in such a link. Systems with
fixed training sequences, such as Wi-Fi, keep the same pattern in
deployment, but there it depends on the amplifier and operating point of
each transmitter, which the receiver does not know
(Section~\ref{sec:disc_appl}).

In the studies we examined, the pilot sequence is not among the
quantities reported as varied between training and testing
(Section~\ref{sec:related}). A fixed
pilot is, however, the default in the open-source Sionna simulator
\citep{hoydis2022sionna}, and the neural-receiver tutorial of that
library depends on it (Section~\ref{sec:res_psn}). Three questions are
therefore open. Does a learned receiver trained with one fixed pilot gain
more on that pilot than on others? Does this also happen for a neural
receiver of an established design, and does it survive long training? And
how large are the gains of learned receivers compared with simple channel
estimates that need no training?

We answer these questions by simulating two- and four-antenna MIMO-OFDM
links with a strongly driven amplifier and oscillator phase noise. We
test two learned receivers, each on the pilot sequence it was trained
with and on sequences it never saw. The first is a network of our own
design with separate weights for every subcarrier. It can store a pattern
across subcarriers by construction, which makes it a positive control for
the test. The second is a convolutional neural receiver with the
architecture of the Sionna tutorial \citep{sionnatutorial}, whose inputs
we extended so that it can be used with any pilot sequence. We also
compare both with two channel estimates that need no training: a
smoothed estimate and an estimate that knows the amplifier.

With binary pilots, both learned receivers gained more on their training
pilot than on unseen pilots. For the per-subcarrier network the difference was about
27 percentage points at the highest signal-to-noise ratio (SNR) tested,
and it grew with longer training. For the neural receiver it was 37
points at the highest SNR after short training and 9 points after nine
times longer training (3 to 8 points at lower SNRs). It remained positive
in every training run, but at the highest SNR much of it then also
appeared with a linear amplifier. Smoothing the channel estimate across
neighbouring subcarriers, which needs neither training nor knowledge of
the amplifier, removed most of the pilot error and lowered the bit error
rate (BER) on every pilot sequence tested, by about 30\% on average at the
highest SNR, as long as the channel's delay spread was short. Of the
learned receivers, only the neural receiver after long training did
better than smoothing on pilots it had not been trained on, and only at
high SNR.

The contributions of this paper are as follows.
\begin{enumerate}
\item An analysis of the pilot error. For a fixed pilot, the amplifier
leaves a deterministic, sequence-specific ripple in the channel estimate
whose size matches the error vector magnitude (EVM) of the data to within
10\%, so it can be predicted from a standard transmitter measurement. It depends on the
pilot family: for quadrature phase-shift keying (QPSK) pilots it also
affects the phase, and it vanishes
for constant-envelope Zadoff--Chu pilots at critical sampling.
\item Evidence that learned receivers memorise this ripple, from a
per-subcarrier network and a training-free model of memorisation tested
on 106 unseen sequences, with controls that remove the pilot distortion;
and the finding that a neural receiver of established architecture also
gains more on its training pilot, by an amount that shrinks with longer
training and is then only partly caused by the amplifier.
\item A test protocol for learned receivers: unseen pilot sequences, a
control with undistorted pilots, and comparison with training-free
estimates, together with the limits of those estimates at long delay
spreads and with an inexact amplifier model.
\end{enumerate}
Section~\ref{sec:related} reviews related work,
Section~\ref{sec:system} describes the link and the receivers,
Section~\ref{sec:protocol} the protocol and the simulations, and
Section~\ref{sec:results} the results. Section~\ref{sec:discussion}
discusses what they mean for the evaluation of learned receivers and
states the limitations, and Section~\ref{sec:conclusion} concludes.

\section{Related work}\label{sec:related}

\begin{table*}[!t]
\caption{Comparison with related studies. }
\label{tab:related}
\footnotesize
\setlength{\tabcolsep}{4pt}
\begin{tabular*}{\textwidth}{@{\extracolsep{\fill}}Y{0.15\textwidth}Y{0.12\textwidth}Y{0.22\textwidth}Y{0.10\textwidth}Y{0.15\textwidth}Y{0.18\textwidth}@{}}
\toprule
Study & PA modelled & Varied between training and testing & Pilot sequence varied & Training-free estimate as reference & Main contribution \\
\midrule
\citet{xu2021} & yes & not stated & not stated & LS mismatch noted, left to the network & model-aided MIMO-OFDM receiver \\
\citet{pihlajasalo2023} & yes, measured PA & PA models, channel profiles, SNR, Doppler & not stated & not stated & convolutional receivers for power efficiency and coverage \\
\citet{nielsen2023} & yes, several PAs & PA models, channels & not stated & not stated & hybrid model training for nonlinear transmitters \\
\citet{aoudia2022} & no & channels & not stated & not applicable & neural receivers and pilotless OFDM \\
\citet{obeed2026}$^{\ast}$ & no & not stated & not stated & not applicable & continual learning from reference signals \\
\citet{kumar2026} & yes, nonlinear distortion & PAs and channels & not stated & not stated & hybrid polar and frequency-domain receiver \\
This work & yes, Rapp and Saleh & pilot sequence & yes & smoothed LS, PA-aware LS & test protocol, pilot fingerprint \\
\bottomrule
\end{tabular*}
\par\smallskip{\footnotesize $^{\ast}$Preprint.}
\end{table*}

Three lines of work are relevant (Table~\ref{tab:related}). The first
evaluates learned receivers under hardware impairments and checks
whether they keep working when the PA model, the channel profile, the impairment level, the number of users or the
modulation changes
\citep{pihlajasalo2022spawc,pihlajasalo2023,nielsen2023,cammerer2023,kumar2026}.
\citet{pihlajasalo2023}, for instance, train on 30 dithered variants of
a measured PA and validate on 10 others. In the studies we examined, the
pilot sequence is not among the quantities reported as changed.

The second line concerns channel estimation with distorted pilots. PA
distortion degrades pilot-based estimation \citep{ligata2011}, classical
receivers re-estimate the channel with a model of the distortion
\citep{sterba2010}, blind estimators avoid pilots \citep{pasangi2019}, and
\citet{xu2021} leave the mismatch of the LS estimate to a neural network. The pilot error is the in-band distortion of a
deterministic signal \citep{dardari2000,banelli2000}, and in models of
transmit impairments the distortion caps the pilot signal-to-noise ratio
at the signal-to-distortion ratio \citep{studer2010,bjornson2014,mollen2016},
so its size is set by the EVM of the transmitter. The LS estimate is also
commonly smoothed across subcarriers \citep{vandebeek1995,edfors1998},
which averages out part of an error that changes from one subcarrier to
the next, and constant-envelope Zadoff--Chu pilots \citep{chu1972}, used
for the low peak-to-average-power uplink reference signals of NR
\citep{3gpp38211}, avoid the error at the source. In the studies we
examined, neither PA-aware nor smoothed estimates are used as the
reference for a learned receiver under PA nonlinearity.

The third line concerns the pilots seen in training. Some neural
receivers take the pilot pattern as an input \citep{aoudia2022} or LS
channel estimates computed from the reference signals
\citep{cammerer2023}, a recent study adapts a neural receiver online from
the reference signals it receives \citep{obeed2026}, and randomising a
nuisance factor during training is a form of domain randomisation
\citep{tobin2017}; none of these considers a nonlinear PA. Testing on
unseen pilots is a grouped validation by pilot sequence and a stress
test in the sense of \citet{damour2022}. Recent studies warn that learned
physical-layer models can lose their benefit outside their training
conditions \citep{xu2026review,elnashar2026}.

\section{System model and receivers}\label{sec:system}

\subsection{Link and power amplifier}\label{sec:link}

We consider an uncoded $N_T\times N_T$ MIMO-OFDM link with
$N_T\in\{2,4\}$, $N_F=64$ subcarriers indexed by $k$, 14 OFDM symbols per
frame indexed by $l$, and Gray-mapped unit-energy 16-point quadrature
amplitude modulation (16-QAM) data
(Table~\ref{tab:params}). Pilots are time-multiplexed: antenna $t$ sends
the same sequence $\mathbf{q}=[q_1,\dots,q_{N_F}]$ on two dedicated OFDM
symbols while the other antennas are silent, namely symbols $\{0,9\}$ and
$\{4,13\}$ at $2\times2$ and $\{t-1, t+7\}$ for antenna
$t\in\{1,\dots,4\}$ at $4\times4$, which leaves 10 and 6 data symbols per
subcarrier. The pilot is a random $\pm1$ (binary phase-shift keying, BPSK)
sequence named after the
integer seed of the NumPy generator that draws it
(\texttt{default\_rng(seed).choice([1,-1], 64)}; seeds 42, 123, 7 and
2024). A ``random'' pilot is a new sequence for every test batch of 64 or
128 frames, which gives 4 to 8 distinct sequences per seed and SNR.
QPSK and Zadoff--Chu pilots are used in Section~\ref{sec:res_pilot}.

Each transmit antenna has its own PA. An OFDM symbol is converted to the
time domain, scaled by $\kappa=10^{-\mathrm{IBO}/20}$, where IBO is the
input back-off in dB, amplified, scaled back and converted back to the
frequency domain:
\begin{equation}
\tilde{\mathbf{x}}=\mathcal{T}(\mathbf{x})
=\mathbf{F}\,\mathcal{P}_\kappa\!\left(\mathbf{F}^{H}\mathbf{x}\right),
\qquad
\mathcal{P}_\kappa(u)=\kappa^{-1}\mathcal{P}(\kappa u),
\label{eq:pa_chain}
\end{equation}
where $\mathbf{F}$ is the unitary discrete Fourier transform (DFT) matrix.
The cyclic prefix does not
change this, because the PA is memoryless. The PA acts on the $N_F$
critically sampled time samples with all 64 subcarriers occupied; a
$4\times$ oversampled PA with guard subcarriers reduces the pilot error
by about a fifth but does not remove it (Section~\ref{sec:res_pilot}).
The PA is $\mathcal{P}(u)=g(|u|)\,e^{\,j\Phi(|u|)}\,u$, either the Rapp
model \citep{rapp1991} with an amplitude-to-phase (AM/PM) conversion term
of our own choice (0.05~rad,
about $3^\circ$, at saturation),
\begin{equation}
g(\rho)=\left[1+\left(\rho/A\right)^{2\nu}\right]^{-1/(2\nu)},\qquad
\Phi(\rho)=\psi\,\frac{\rho^{2}}{A^{2}+\rho^{2}},
\label{eq:rapp}
\end{equation}
with $A=1$, $\nu=2$ and $\psi=0.1$, or the Saleh model \citep{saleh1981}
\begin{equation}
g(\rho)=\frac{1}{1+c_1\rho^{2}},\qquad
\Phi(\rho)=\frac{c_2\,\rho^{2}}{1+c_3\,\rho^{2}},
\label{eq:saleh}
\end{equation}
with $c_1=1.1517$, $c_2=4.0033$, $c_3=9.1040$ and unit small-signal gain.
The Saleh PA compresses gradually, rotates the phase by up to 0.44~rad and
saturates at a lower input amplitude ($1/\sqrt{c_1}\approx0.93$), so the
same IBO does not mean the same compression. Neither PA is linearised by
predistortion. Unless stated otherwise, the PA is the Rapp model at
$\mathrm{IBO}=3$~dB.

\begin{figure*}[t]
\centering
\begin{tikzpicture}[
  font=\small, node distance=4mm and 4.5mm,
  blk/.style={draw, rounded corners=2pt, minimum height=9mm,
              align=center, text width=17mm, fill=white},
  rxb/.style={blk, fill=black!6},
  key/.style={blk, fill=black!6, very thick},
  arr/.style={-{Stealth[length=2.2mm]}, semithick}]
\node[blk] (src) {16-QAM data and pilot $\mathbf{q}$};
\node[blk, right=of src] (ifft) {IFFT and cyclic prefix};
\node[blk, right=of ifft] (pa) {PA $\mathcal{P}_\kappa(\cdot)$};
\node[blk, right=of pa] (pn) {Phase noise $e^{\,j\varphi_t(n)}$};
\node[blk, right=of pn] (ch) {Channel $\mathbf{H}_k$ and noise};
\node[rxb, below=8mm of ch] (fft) {Remove prefix and FFT};
\node[key, left=of fft] (ls) {LS estimate $y/r_k$, optional smoothing};
\node[rxb, left=of ls] (eq) {Interpolation and LMMSE};
\node[rxb, left=of eq] (trk) {Phase tracker};
\node[rxb, left=of trk] (lrn) {Learned stage (optional)};
\node[rxb, left=of lrn] (dec) {Decisions and BER};
\node[blk, below=6mm of ls, text width=58mm] (ref)
  {Reference $r_k$: plain LS $r_k=q_k$, or PA-aware LS $r_k=\tilde q_k/\alpha$; smoothing by Eq.~\eqref{eq:smooth}};
\node[rxb, below=6mm of lrn, text width=36mm, dashed] (nrx)
  {Neural receiver: received grid, LS estimate and LMMSE output $\rightarrow$ bit LLRs};
\draw[arr] (src)--(ifft); \draw[arr] (ifft)--(pa); \draw[arr] (pa)--(pn);
\draw[arr] (pn)--(ch); \draw[arr] (ch)--(fft); \draw[arr] (fft)--(ls);
\draw[arr] (ls)--(eq); \draw[arr] (eq)--(trk); \draw[arr] (trk)--(lrn);
\draw[arr] (lrn)--(dec); \draw[arr] (ref)--(ls);
\draw[arr, dashed] (trk.south) |- (nrx.east);
\draw[arr, dashed] (nrx.west) -| (dec.south);
\end{tikzpicture}
\caption{Simulated MIMO-OFDM link and receivers .}
\label{fig:pipeline}
\end{figure*}
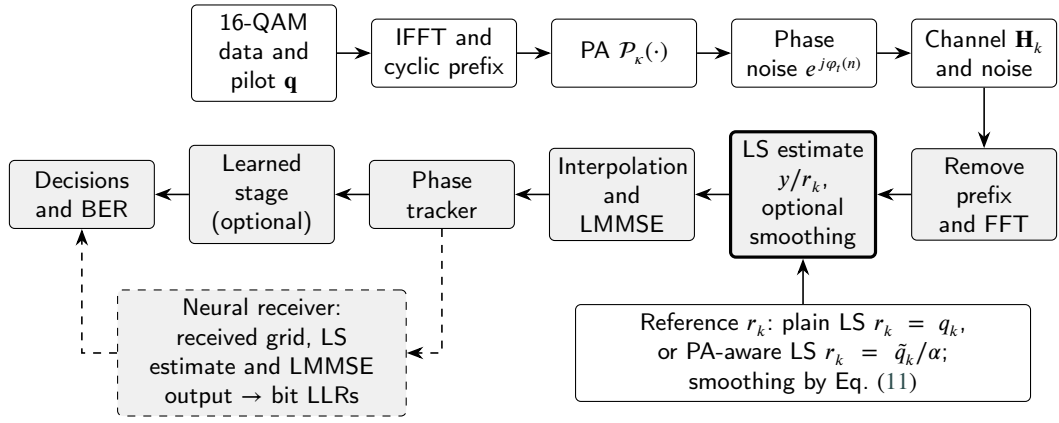

After the PA, the signal of antenna $t$ is multiplied by
$e^{\,j\varphi_t(n)}$, where $n$ is the time-sample index and
$\varphi_t(n)$ an independent Wiener phase noise with a linewidth of
100~Hz at the sampling rate of 1.92~MHz \citep{wu2004}. The received
signal on subcarrier $k$ of OFDM symbol $l$ is
\begin{equation}
\mathbf{y}[l,k]=\mathbf{H}_k\,\check{\mathbf{x}}[l,k]+\mathbf{n}[l,k],
\label{eq:rx}
\end{equation}
where $\mathbf{H}_k$ is the $N_T\times N_T$ channel matrix,
$\check{\mathbf{x}}[l,k]$ collects the amplified, phase-rotated symbols
of all antennas, and
$\mathbf{n}[l,k]\sim\mathcal{CN}(\mathbf{0},N_0\mathbf{I})$, with $N_0$
set by the ratio $E_b/N_0$ of the energy per bit to the noise spectral
density, for 4 bits per symbol at the PA input. We report this
$E_b/N_0$ as the SNR of a result. The PA output power is lower, by 1.3~dB (Rapp) and 5.2~dB (Saleh) at
$\mathrm{IBO}=3$~dB. The channel is the tapped-delay-line profile TDL-A of
the 3rd Generation Partnership Project (3GPP) with 30~ns delay spread
\citep{3gpp38901} or, in some runs, independent and identically
distributed (i.i.d.)\ Rayleigh fading that is flat across subcarriers;
the smoothing and delay-spread studies also use TDL-C at
100~ns and both profiles at 300 to 1000~ns. The channel is constant over a
frame and applied in the frequency domain, so inter-symbol interference
is ignored. The EVM and adjacent-channel limits quoted later are those of
a base station \citep{3gpp38104}. Fig.~\ref{fig:pipeline} summarises the
chain.

\subsection{What the PA does to data and to pilots}\label{sec:pilotpa}

For data symbols, the Bussgang decomposition
\citep{bussgang1952,demir2021bussgang} splits the PA output on each
subcarrier into a scaled input and an uncorrelated distortion,
\begin{equation}
\tilde x_k=\alpha\,x_k+d_k,\qquad
\alpha=\frac{\mathbb{E}\{\tilde x_k\,x_k^{*}\}}{\mathbb{E}\{|x_k|^{2}\}}.
\label{eq:bussgang}
\end{equation}
Because the time samples of an OFDM symbol carrying random data are close
to Gaussian, $\alpha$ is one complex number for all subcarriers and
antennas: $\alpha=0.849+j0.036$ (Rapp) and $0.490+j0.184$ (Saleh) at
$\mathrm{IBO}=3$~dB. A pilot symbol is different. The distorted pilot
$\tilde{\mathbf{q}}=\mathcal{T}(\mathbf{q})$ is deterministic and the same
in every frame. We describe it by the gain that the PA applies to each
pilot symbol,
\begin{equation}
b_k=\frac{\tilde q_k}{q_k}=\alpha+\frac{d_k}{q_k},
\label{eq:bk}
\end{equation}
which depends on the whole sequence. For a deterministic pilot this
equation defines $d_k$; it is not an application of Bussgang's theorem.
The ripple of $b_k$ across subcarriers is the in-band distortion of the
pilot \citep{dardari2000,banelli2000}, fixed from frame to frame, and we
call this fixed pattern the \emph{pilot fingerprint}. Throughout, the data
EVM is the rms of the distortion (output minus $\alpha$ times input)
relative to the rms of $\alpha$ times the input, measured without
equalisation and, unless stated, at critical sampling (so not with the
3GPP test procedure). The ripple is the rms deviation of $b_k/\alpha$ from
its mean over subcarriers.

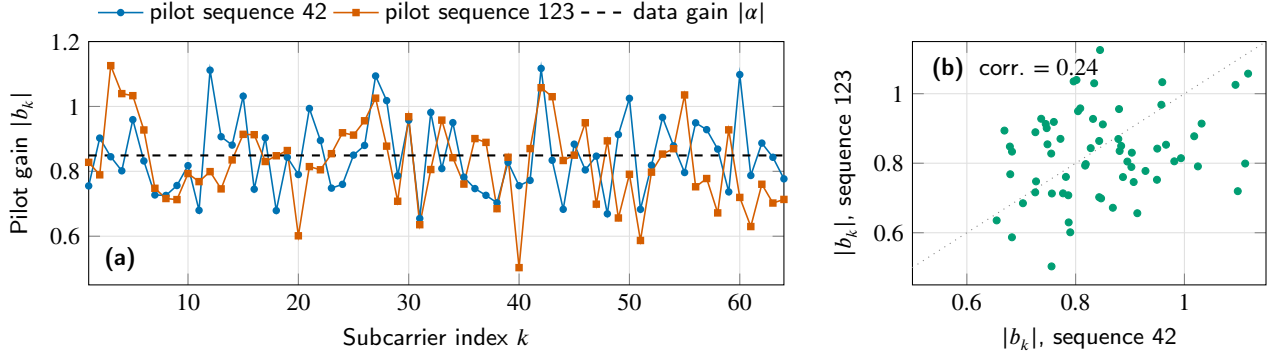
\begin{figure*}[t]
\centering
\begin{tikzpicture}
\begin{axis}[name=pa,
  width=0.62\textwidth, height=48mm,
  xmin=1, xmax=64, ymin=0.45, ymax=1.2,
  xlabel={Subcarrier index $k$}, ylabel={Pilot gain $|b_k|$},
  grid=major, grid style={black!12},
  legend style={at={(0.0,1.03)}, anchor=south west, draw=none,
                fill=none, legend columns=3, font=\small},
  tick label style={font=\small}, label style={font=\small}]
\addplot[color=cbblue, mark=*, mark size=1.1pt, line width=0.6pt]
  coordinates {(1,0.7551) (2,0.9025) (3,0.8449) (4,0.8016) (5,0.9593) (6,0.8320) (7,0.7272) (8,0.7257) (9,0.7563) (10,0.8175) (11,0.6798) (12,1.1117) (13,0.9067) (14,0.8809) (15,1.0316) (16,0.7450) (17,0.9033) (18,0.6791) (19,0.8433) (20,0.7901) (21,0.9937) (22,0.8951) (23,0.7482) (24,0.7599) (25,0.8500) (26,0.8798) (27,1.0939) (28,1.0180) (29,0.7865) (30,0.9576) (31,0.6548) (32,0.9814) (33,0.8085) (34,0.9501) (35,0.7821) (36,0.7468) (37,0.7260) (38,0.7031) (39,0.8277) (40,0.7557) (41,0.7721) (42,1.1171) (43,0.8341) (44,0.6830) (45,0.8837) (46,0.8044) (47,0.8468) (48,0.6691) (49,0.9134) (50,1.0248) (51,0.6829) (52,0.8186) (53,0.9662) (54,0.8785) (55,0.7963) (56,0.9495) (57,0.9284) (58,0.8684) (59,0.7367) (60,1.0981) (61,0.7872) (62,0.8870) (63,0.8433) (64,0.7768)};
\addlegendentry{pilot sequence 42}
\addplot[color=cbverm, mark=square*, mark size=1.1pt, line width=0.6pt]
  coordinates {(1,0.8278) (2,0.7892) (3,1.1257) (4,1.0395) (5,1.0332) (6,0.9275) (7,0.7477) (8,0.7162) (9,0.7129) (10,0.7931) (11,0.7683) (12,0.7991) (13,0.7459) (14,0.8352) (15,0.9140) (16,0.9129) (17,0.8303) (18,0.8480) (19,0.8641) (20,0.6013) (21,0.8145) (22,0.8048) (23,0.8544) (24,0.9188) (25,0.9117) (26,0.9558) (27,1.0255) (28,0.8776) (29,0.7078) (30,0.9682) (31,0.6355) (32,0.8055) (33,0.9578) (34,0.8420) (35,0.7607) (36,0.9009) (37,0.8892) (38,0.6849) (39,0.8434) (40,0.5032) (41,0.8701) (42,1.0577) (43,1.0300) (44,0.8334) (45,0.8498) (46,0.9497) (47,0.6988) (48,0.8939) (49,0.6562) (50,0.7909) (51,0.5869) (52,0.7975) (53,0.8532) (54,0.8697) (55,1.0354) (56,0.7524) (57,0.7778) (58,0.6719) (59,0.9279) (60,0.7196) (61,0.6298) (62,0.7602) (63,0.7023) (64,0.7134)};
\addlegendentry{pilot sequence 123}
\addplot[black, dashed, line width=0.8pt, domain=1:64, samples=2] {0.849};
\addlegendentry{data gain $|\alpha|$}
\node[anchor=south west, font=\small\bfseries] at (rel axis cs:0.01,0.02) {(a)};
\end{axis}
\begin{axis}[at={(pa.south east)}, xshift=17mm, anchor=south west,
  width=0.36\textwidth, height=48mm,
  xmin=0.5, xmax=1.15, ymin=0.45, ymax=1.15,
  xlabel={$|b_k|$, sequence 42}, ylabel={$|b_k|$, sequence 123},
  grid=major, grid style={black!12},
  tick label style={font=\small}, label style={font=\small}]
\addplot[only marks, mark=*, mark size=1.3pt, color=cbgreen]
  coordinates {(0.7551,0.8278) (0.9025,0.7892) (0.8449,1.1257) (0.8016,1.0395) (0.9593,1.0332) (0.8320,0.9275) (0.7272,0.7477) (0.7257,0.7162) (0.7563,0.7129) (0.8175,0.7931) (0.6798,0.7683) (1.1117,0.7991) (0.9067,0.7459) (0.8809,0.8352) (1.0316,0.9140) (0.7450,0.9129) (0.9033,0.8303) (0.6791,0.8480) (0.8433,0.8641) (0.7901,0.6013) (0.9937,0.8145) (0.8951,0.8048) (0.7482,0.8544) (0.7599,0.9188) (0.8500,0.9117) (0.8798,0.9558) (1.0939,1.0255) (1.0180,0.8776) (0.7865,0.7078) (0.9576,0.9682) (0.6548,0.6355) (0.9814,0.8055) (0.8085,0.9578) (0.9501,0.8420) (0.7821,0.7607) (0.7468,0.9009) (0.7260,0.8892) (0.7031,0.6849) (0.8277,0.8434) (0.7557,0.5032) (0.7721,0.8701) (1.1171,1.0577) (0.8341,1.0300) (0.6830,0.8334) (0.8837,0.8498) (0.8044,0.9497) (0.8468,0.6988) (0.6691,0.8939) (0.9134,0.6562) (1.0248,0.7909) (0.6829,0.5869) (0.8186,0.7975) (0.9662,0.8532) (0.8785,0.8697) (0.7963,1.0354) (0.9495,0.7524) (0.9284,0.7778) (0.8684,0.6719) (0.7367,0.9279) (1.0981,0.7196) (0.7872,0.6298) (0.8870,0.7602) (0.8433,0.7023) (0.7768,0.7134)};
\addplot[black!50, dotted, domain=0.45:1.15, samples=2] {x};
\node[anchor=north west, font=\small] at (rel axis cs:0.03,0.97) {\textbf{(b)}\; corr.\ $=0.24$};
\end{axis}
\end{tikzpicture}
\caption{Pilot gain $|b_k|$ for pilot sequences 42 and 123 : (a) per subcarrier, (b) sequence 123 against sequence 42.}
\label{fig:bias}
\end{figure*}

The time signal of a random BPSK or QPSK pilot is close to Gaussian, so
$d_k/q_k$ behaves like one realisation of the data distortion. The ripple
therefore approximately equals the data EVM, and the mismatch
$b_k(\mathbf{q})/b_k(\mathbf{q}')$ between two sequences deviates from one
by about $\sqrt2$ times the EVM, which would be exact for independent
ripples. For a real $\pm1$ pilot the time signal is conjugate-symmetric,
which a PA with weak AM/PM nearly preserves, so the ripple lies in the
amplitude of $b_k$; for QPSK pilots it also has a phase part.
Fig.~\ref{fig:bias} shows $|b_k|$ for two sequences (Rapp,
$\mathrm{IBO}=3$~dB, no noise). Its mean (0.85 and 0.83) is close to
$|\alpha|=0.85$, but it varies across subcarriers with a standard
deviation of 0.115 and 0.125 (14 and 15\% of the mean), and the two
patterns are only weakly related (correlation 0.24). The data EVM at this
operating point is 13.7\%, and the ripple over 200 random sequences is
13.8\%.

The plain LS estimator divides by the ideal pilot, $r_k=q_k$. Without
noise and phase noise it gives
$\widehat{\mathbf{H}}_k^{\mathrm{LS}}=b_k\,\mathbf{H}_k$ for every antenna
pair, because all antennas send the same sequence through identical PAs.
The data, however, see $\alpha\mathbf{H}_k$, so after equalisation every
data symbol on subcarrier $k$ is multiplied by
\begin{equation}
\mu_k=\frac{\alpha}{b_k},
\label{eq:mu}
\end{equation}
close to one on average but about 14\% off on a typical subcarrier, in a
pattern fixed for a given sequence. This is the degradation analysed by
\citet{ligata2011} and the mismatch noted by \citet{xu2021}.
Eq.~\eqref{eq:mu} neglects the regularisation of the linear minimum
mean-square error (LMMSE) equaliser; in
a Monte Carlo check with Rayleigh channels and the estimated channel
scaled by $|\alpha|$, the median shrinkage of the weakest eigen-direction
is 1.5\% ($2\times2$) and 3.1\% ($4\times4$) at 18~dB and 0.10 and 0.20\%
at 30~dB. The PA-aware LS estimator divides by the pilot as it leaves the
PA, scaled by the data gain,
\begin{equation}
r_k=\frac{\tilde q_k}{\alpha},\qquad
\widehat{\mathbf{H}}_k^{\mathrm{PA}}=\alpha\,\mathbf{H}_k ,
\label{eq:pals}
\end{equation}
which is exactly the channel seen by the data. It needs no training and
works for any pilot sequence, but the receiver must know the PA model and
the back-off. Re-estimating the channel with a model of the nonlinearity
is classical \citep{sterba2010}. We use Eq.~\eqref{eq:pals} as an oracle
reference, and Section~\ref{sec:res_limits} studies wrong PA assumptions
and a reference measured by a loopback.

\subsection{Training-free receivers}\label{sec:chain}

The baseline receiver, called the LMMSE baseline, interpolates and
extrapolates the plain LS estimates of each antenna's two pilot symbols
linearly in time and equalises \citep{tse2005}
\begin{equation}
\hat{\mathbf{s}}[l,k]=\left(\widehat{\mathbf{H}}^{H}\widehat{\mathbf{H}}
+N_0\mathbf{I}\right)^{-1}\widehat{\mathbf{H}}^{H}\mathbf{y}[l,k],
\label{eq:lmmse}
\end{equation}
with the interpolated estimate $\widehat{\mathbf{H}}$. It then removes the
common phase error of each stream $t$ and OFDM symbol $l$ with a
decision-directed tracker following \citet{wu2004},
\begin{equation}
\hat\theta_t[l]=\angle\sum_{k=1}^{N_F}\hat s_t[l,k]\,\hat c_t^{*}[l,k],\qquad
\bar s_t[l,k]=\hat s_t[l,k]\,e^{-j\hat\theta_t[l]},
\label{eq:tracker}
\end{equation}
where $\hat c_t[l,k]$ is the nearest 16-QAM point, repeated three times
with recomputed decisions. If the distortion is modelled as white on each
stream and passed through the channel, a distortion-aware LMMSE equaliser
(cf.\ \citet{demir2021bussgang}) replaces $N_0$ by $N_0/(1+\mathrm{EVM}^2)$.
At $\mathrm{IBO}=3$~dB this changes the regularisation by 1.8\% and gives
the same BER (0.02456 at 30~dB, pilot 42), whereas adding the distortion
to $N_0$ as white noise raises the BER to 0.0455, so
Eq.~\eqref{eq:lmmse} is not a weakened baseline. The smoothed receiver
filters the LS estimate across subcarriers before the equaliser,
\begin{equation}
\widehat{\mathbf{H}}_k^{\mathrm{SM}}=\sum_{k'=1}^{N_F}s_{kk'}\,
\widehat{\mathbf{H}}_{k'}^{\mathrm{LS}},
\label{eq:smooth}
\end{equation}
where the weights $s_{kk'}$ are the elements of a smoothing matrix
$\mathbf{S}$; we use two standard choices \citep{vandebeek1995,edfors1998}.
The local linear smoother LIN-$M$ fits a straight line by least squares
to the $M$ subcarriers centred on $k$, with the window clipped at the band
edges. The LMMSE smoother LM-$\tau$ uses
$\mathbf{S}=\boldsymbol{\Sigma}(\boldsymbol{\Sigma}+\sigma_s^2\mathbf{I})^{-1}$,
where $\Sigma_{kk'}=1/\bigl(1+j2\pi\tau f_{\mathrm{sc}}(k-k')\bigr)$ is
the frequency correlation of an exponential power-delay profile with rms
delay $\tau$, $f_{\mathrm{sc}}=30$~kHz, and $\sigma_s^2=0.02$ is the
assumed relative error variance of the LS estimate. LIN-9 and
$\sigma_s^2=0.02$ were fixed before the experiments. Neither smoother is
trained or uses knowledge of the PA. Because $b_k$ varies between
neighbouring subcarriers while $\mathbf{H}_k$ varies slowly, smoothing
averages $b_k$ towards its mean. Hard decisions on
$\bar{\mathbf{s}}[l,k]$, or on the output of a learned stage, give the
BER.

\subsection{Learned receivers}\label{sec:learned}

\subsubsection{Learned stages after the tracker}

A learned stage maps the
tracked data symbols of a frame to refined symbols. We compare four
stages (numbers of real parameters at $2\times2$ and $4\times4$ in
parentheses): a complex gain per stream (4, 8); a gain and cubic term
$\mathbf{s}'=\mathbf{M}\bar{\mathbf{s}}
+\mathbf{C}\,(\bar{\mathbf{s}}\odot|\bar{\mathbf{s}}|^{2})$ with matrices
$\mathbf{M},\mathbf{C}$ shared by all subcarriers and element-wise
$\odot$ and $|\cdot|^2$ (16, 64); the per-subcarrier network (PSN;
11,310 and 27,650); and the shared-matrix network (SMN; 8,790 and
17,570). The PSN has five layers in cascade. Layer $i$ applies the linear
map
$(\mathbf{B}^{(i)H}_k\mathbf{B}^{(i)}_k+\sigma_i^{2}\mathbf{I})^{-1}\mathbf{B}^{(i)H}_k$,
built from a learned $N_T\times N_T$ complex matrix $\mathbf{B}^{(i)}_k$
for every subcarrier and a learned $\sigma_i^2$, and adds a learned
multiple of the output of a two-layer convolutional network
(3$\times$3 kernels, 16 hidden channels) over the time-frequency grid.
The SMN is the same network with one matrix $\mathbf{B}^{(i)}$ shared by
all subcarriers. We designed the PSN as a stand-in for receivers with
subcarrier-specific parameters and as a positive control for the test
protocol; it is not a model of a published receiver.

The stages are trained with the binary cross-entropy of the bit
log-likelihood ratios (LLRs) computed with the thermal-noise variance
$N_0$, for 500 iterations, each on a batch of newly generated frames
(5,000 iterations in the training-length study), with the settings of
Table~\ref{tab:params}.
Fixed-pilot training uses sequence 42, random-pilot training a new
sequence in every batch. Stages after the PA-aware estimate are trained
with random pilots, stages after a smoothed estimate with sequence 42.
The hyperparameters were set during development with one fixed sequence
(123 in the earliest runs, 42 later) and not re-tuned. Sequences 7 and
2024 played no role in the design, and the PSN loses almost equally on
123 and 7 ($-2.6$ and $-2.7\%$ at 30~dB, $2\times2$, TDL-A, ten seeds),
so the loss is not an artefact of development.

A stage receives, for every data resource element,
\begin{equation}
\bar{\mathbf{s}}[l,k]\approx\mu_k\,\mathbf{x}[l,k]+\mathbf{e}[l,k],
\label{eq:model_in}
\end{equation}
where $\mathbf{e}[l,k]$ collects PA distortion, residual phase noise,
estimation noise and thermal noise ($\mu_k=1$ with the PA-aware
estimate). A lower BER can come from undoing $\mu_k$, an error of the
channel estimate that depends on the pilot sequence, or from reducing
$\mathbf{e}[l,k]$, which is true compensation of the PA. With fixed-pilot
training $\mu_k$ is the same in every frame, and the PSN cancels it
exactly with $\mathbf{B}^{(i)}_k=\mu_k\mathbf{I}$, whereas a shared map
can correct only the mean of $\mu_k$, which is close to one.

\subsubsection{Neural receiver}

To test a receiver of established design, we
ported the neural receiver of the Sionna tutorial \citep{sionnatutorial},
which follows \citet{honkala2021deeprx} and \citet{aoudia2022}, to
PyTorch. It has a $3\times3$ input convolution with 128 channels, four
residual blocks, each with two layer normalisations, rectified linear unit
(ReLU) activations and
two $3\times3$ convolutions with 128 channels, and a $3\times3$ output
convolution with one channel per bit (eight at $2\times2$ with 16-QAM),
and it processes the whole $14\times64$ resource grid at once. As in the
tutorial, each layer normalisation is computed jointly over the OFDM
symbols, subcarriers and channels and has a learned scale and offset for
every element. The convolutions share their weights across the grid, but
1.8 of the network's 3.0 million parameters are tied to individual
resource elements.
The tutorial network receives only the received grid and the noise
variance, so it relies on the pilots being the same as in training. So that it can
be tested on other pilots, we also give it the plain LS estimate and the
output of the LMMSE equaliser and tracker (17 input channels at
$2\times2$: real and imaginary parts of the two received signals, the
four channel coefficients and the two equalised streams, and
$\log_{10}N_0$); the multi-user receiver of \citet{cammerer2023} likewise
takes interpolated LS estimates as an input. We refer to this network as
the neural receiver (NRX). It is trained with the binary cross-entropy of
its bit outputs on the data resource elements, batches of 32 frames,
$E_b/N_0$ drawn uniformly from 10 to 30~dB, the Adam optimiser and a
cosine learning-rate schedule from $10^{-3}$ to $10^{-5}$, for 4,000,
12,000 or 36,000 iterations. As for the stages, fixed-pilot training uses
sequence 42 and random-pilot training a new sequence in every batch.
During training we record the loss at
24~dB on a fixed validation batch sent with the training pilot and with a
held-out pilot (sequence 9999, not used for testing). The tutorial trains
for 30,000 iterations with batches of 128 frames, so our longest runs saw
about 30\% of its training frames; Section~\ref{sec:res_nrx} shows that
the validation loss had levelled off by then, although this does not
prove convergence.

\subsection{A training-free model of memorisation}\label{sec:memoriser}

Complete memorisation can be modelled without training. The
\emph{memoriser} of a fixed pilot $\mathbf{q}$ divides the received pilot
$\mathbf{q}'$ by $r_k=q'_k\,b_k(\mathbf{q})/\alpha$. On $\mathbf{q}$ it
equals the PA-aware estimate; on $\mathbf{q}'$ it leaves the factor
$b_k(\mathbf{q}')/b_k(\mathbf{q})$. Two variants store only the amplitude
or only the phase of $b_k(\mathbf{q})/\alpha$. The memoriser reproduces
the gain of the trained PSN on its training pilot, to within 1 percentage
point (pp) at 24 and 30~dB, and loses more on unseen pilots
(Section~\ref{sec:res_psn}), so it is a pessimistic, training-free
stand-in for a memorising stage; it lost more than the PSN in every BPSK
and QPSK case we compared, but we did not prove that this always holds. Performance is
measured by the BER reduction relative to the LMMSE baseline on the same
frames and pilot sequence,
\begin{equation}
R=100\,\frac{\mathrm{BER}_0-\mathrm{BER}}{\mathrm{BER}_0}\ \%.
\label{eq:R}
\end{equation}

\section{Evaluation protocol and simulation setup}\label{sec:protocol}

The link is simulated with Sionna 1.2.2 \citep{hoydis2022sionna} and the
learned receivers in PyTorch (Table~\ref{tab:params}). Except in the
coded runs of Section~\ref{sec:res_smooth}, the BER is computed by hard
decisions on uncoded 16-QAM symbols at $E_b/N_0$ of 18, 24 and 30~dB;
within one experiment and seed, all methods see the same frames. The protocol has three checks, which apply to any learned
receiver.

\subsection{Check 1: unseen pilot sequences}

A receiver trained with
sequence 42 is tested with it and with unseen sequences (123, 7, 2024 and
random). For each seed we compute the gap
\begin{equation}
D=R_{\mathrm{train}}-R_{\mathrm{unseen}},
\label{eq:drop}
\end{equation}
where $R_{\mathrm{train}}$ is the BER reduction on the training sequence
and $R_{\mathrm{unseen}}$ the mean over the unseen \emph{fixed}
sequences, both relative to the baseline with the same channel estimate
as the receiver's input (plain LS unless stated). Random pilots are
reported separately. A receiver that compensates the PA, and nothing
else, should show $D\approx0$. The memoriser is also tested on 106 unseen
sequences for each of five training pilots, with a bootstrap interval
over sequences.

\subsection{Check 2: controls without pilot distortion}

The PSN and the
SMN are trained and tested with pilots that bypass the PA (clean pilots)
and with the true channel (ideal channel). For the neural receiver, the
whole link is repeated with a linear PA. If an advantage exists only with
PA-distorted pilots, it comes from the pilots. This check needs the
simulator.

\subsection{Check 3: simple references}

We compare each learned receiver with two estimates that need no
training, the smoothed estimate and the PA-aware estimate. The learned
stages are also
trained on top of these estimates and compared there with two minimal
stages, a per-stream gain and a gain with a cubic term. All results are
reported with the absolute BER of the reference.

\begin{table}[t]
\caption{Simulation parameters.}\label{tab:params}
\footnotesize
\begin{tabular}{@{}Y{0.33\columnwidth}Y{0.62\columnwidth}@{}}
\toprule
Parameter & Value \\
\midrule
Antennas & $2\times2$, $4\times4$ \\
Subcarriers, spacing & 64, 30~kHz (sampling rate 1.92~MHz) \\
OFDM symbols per frame, cyclic prefix & 14, 16 samples \\
Carrier frequency & 3.5~GHz \\
Modulation & Gray-mapped 16-QAM, uncoded \\
Pilots & time-multiplexed, two OFDM symbols per antenna (overhead 28.6\% at $2\times2$, 57.1\% at $4\times4$); BPSK from NumPy \texttt{default\_rng(seed)}, or QPSK, Zadoff--Chu \\
Input back-off & 3~dB; 1 to 9~dB in the back-off studies \\
Phase noise & Wiener, 100~Hz linewidth, independent per transmit antenna \\
Channel & 3GPP TDL-A, 30~ns \citep{3gpp38901}, or i.i.d.\ Rayleigh; TDL-C, 100~ns; TDL-A and TDL-C with 300 to 1000~ns in the delay-spread study; constant over a frame, no inter-symbol interference \\
Test $E_b/N_0$ & 18, 24, 30~dB (22 and 30~dB in the coded runs) \\
Training, stages & 500 or 5,000 iterations, Adam, learning rate $10^{-3}\to10^{-5}$ (cosine), batch of 64 ($2\times2$) or 32 ($4\times4$) frames, $E_b/N_0\sim\mathcal{U}[10,25]$~dB \\
Training, neural receiver & 4,000, 12,000 or 36,000 iterations, Adam, learning rate $10^{-3}\to10^{-5}$ (cosine), batch of 32 frames, $E_b/N_0\sim\mathcal{U}[10,30]$~dB \\
Frames per BER value & 1,024 ($2\times2$), 256 ($4\times4$); 512 or 768 in some studies, 192 in the $4\times4$ sensitivity study; 32 or 64 per sequence in the many-pilot studies \\
Software, hardware & Sionna 1.2.2, PyTorch; NVIDIA GPUs (Kaggle) \\
\bottomrule
\end{tabular}
\end{table}

\subsection{Simulation setup and statistics}

Table~\ref{tab:params} gives the simulation parameters, and
Table~\ref{tab:exps} in Appendix~\ref{app:exps} lists the experiments. A
seed is one training run followed by testing. In four experiments of
Table~\ref{tab:exps} (neural receiver, NRX training length, trained
back-off and stage training length), all seeds are tested on the same
frames, so their intervals reflect the variability of training only; in
the other experiments each seed is tested on new frames, so the intervals
also include test-set variability. We report the mean over seeds
with a Student-$t$ 95\% confidence interval; differences between methods
are computed per seed on the same frames, in pp. We do not correct for
multiple comparisons, so the intervals are descriptive. Seed numbers were
fixed before the runs. Experiments with three seeds have wide intervals
(two degrees of freedom), and the neural receiver trained with random
pilots for 36,000 iterations has two seeds, for which we give the
individual values. The conclusions rest on differences of several pp
with the same sign in every seed; sub-pp differences, such as 0.13~pp
between the PSN and the SMN trained on random pilots, are reported as
bounds, not as effects.

\section{Results}\label{sec:results}

Unless stated otherwise, results are for $2\times2$, TDL-A and the Rapp
PA at $\mathrm{IBO}=3$~dB.

\subsection{A per-subcarrier network memorises the pilot}
\label{sec:res_psn}

\begin{figure*}[t]
\centering
\includegraphics[width=\textwidth]{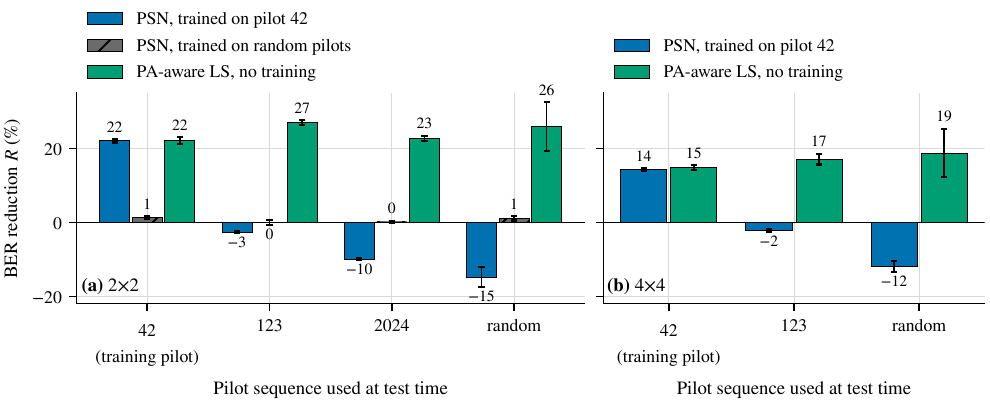}
\caption{BER reduction per test pilot sequence at 30~dB: (a) $2\times2$ , (b) $4\times4$ .}
\label{fig:pilot_test}
\end{figure*}

Fig.~\ref{fig:pilot_test} shows Check~1 for the PSN. Trained and tested
with sequence 42, it lowers the BER of the LMMSE baseline at 30~dB by
22\% ($2\times2$) and 14\% ($4\times4$); with any other sequence it raises
it, by 2 to 15\% (up to 18\% in the runs of Table~\ref{tab:smooth}). In
the 10-seed $2\times2$ run its BER reduction is
$+21.9\%$ on sequence 42 and $-2.6$, $-2.7$, $-9.7$ and $-15.0\%$ on
sequences 123, 7, 2024 and random pilots, which gives $D=26.8$~pp [26.5,
27.1], with a negative unseen-pilot result in all 10 seeds. The i.i.d.\
channel gives the same results within about one pp. In absolute terms the
BER at 30~dB falls from 0.024 to 0.019 on the training sequence but rises
to 0.027 on unseen ones. At $4\times4$, $D=19.2$~pp [18.0, 20.4]
(sequences 123 and 2024, 4 seeds, Table~\ref{tab:smooth}). The gap grows with SNR, from 14.1 to
22.1 and 26.8~pp at 18, 24 and 30~dB, so it is present inside the
training range of 10 to 25~dB.

The gap also grows with training. In a separate three-seed run, 5,000
instead of 500 iterations raised it from 27.2 to 37.2~pp [36.4, 38.0] at
30~dB. Over the
same training the validation loss on the held-out pilot rises (from 0.57
early in training to 0.61 after 500 iterations and 0.69 after 5,000, at
24~dB), while the loss on the training pilot falls from 0.49 to 0.43: the
usual signature of memorisation.

The mechanism follows from Section~\ref{sec:pilotpa}. Trained with one
sequence, the PSN sees the same $\mu_k=\alpha/b_k$ in every frame and
stores its inverse. With another sequence $\mathbf{q}'$ the symbols leave
the network multiplied by $b_k(\mathbf{q})/b_k(\mathbf{q}')$, whose rms
deviation from one (0.16 to 0.20) exceeds that of $\mu_k$ (0.12 to 0.18),
so the BER reduction becomes negative instead of merely vanishing. Three
observations show that the per-subcarrier parameters store the pattern.
The SMN has no comparable gap even with fixed-pilot training (between
$-0.5$ and $1.4\%$ on every sequence, $D=1.5$~pp). Trained with random
pilots, the PSN gains only about 1\%, no more than the SMN. And in a
20-seed run, sharing the PSN's matrices across blocks of 2, 4, 8 and 16
subcarriers lowered its gain from 22.0\% to 12.2, 6.0, 3.7 and 2.1\%,
close to the 1.4\% of a fully shared matrix. This is what we expect if
the network can store only the block average of a nearly white pattern.

The memoriser reproduces the training-pilot gain of the PSN (21.1 against
22.1\% at 30~dB and 18.2 against 18.6\% at 24~dB, 512 frames) and loses
more on unseen pilots ($-12.9$ and $-23.6\%$ on sequences 123 and 2024,
against $-2.6$ and $-9.9\%$ for the PSN in a 5-seed run on other frames):
the trained network stores most,
but not all, of the pattern. Storing only the amplitude of $b_k/\alpha$
gives almost the same result ($+20.6$ and $-26.0\%$ on sequence 42 and on
random pilots), storing only its phase almost nothing ($+1.6$ and
$-0.3\%$), as expected for BPSK pilots. Table~\ref{tab:many} applies
Check~1 with the memoriser to five training pilots, each tested on the
other 106 of 107 sequences. Every training pilot gains (14 to 25\%) and
loses on 92 to 100\% of the unseen sequences, with $D$ between 23 and
86~pp, and the loss grows with the mismatch
$b_k(\mathbf{q}')/b_k(\mathbf{q})$ (pooled correlation $-0.66$). The fixed
test sequences were favourable: for pilot 42, sequences 123, 7 and 2024
lie at the 97th, 94th and 79th percentiles of the 106 unseen results,
which is why random pilots lose more than the mean of the fixed ones.
Smoothing (LIN-9) and the PA-aware estimate, by contrast, help on all 107
sequences, by at least 13.3 and 10.5\% (mean 29.7 and 23.5\%).

\begin{table}[t]
\caption{Check 1 with the memoriser: five training pilots, each tested on 106 unseen sequences.}
\label{tab:many}
\footnotesize
\setlength{\tabcolsep}{1.4pt}
\begin{tabular*}{\columnwidth}{@{\extracolsep{\fill}}rrrrrr@{}}
\toprule
Training & $R_{\mathrm{train}}$ & \multicolumn{2}{c}{$R_{\mathrm{unseen}}$} & Fraction & $D$ \\
\cmidrule(lr){3-4}
pilot & & mean [95\% CI] & 5--95th pct. & $<0$ & (pp) \\
\midrule
42 & 20.4 & $-35.0$ [$-37.3$, $-32.6$] & $-53.7$, $-13.7$ & 1.00 & 55.4 \\
7 & 14.0 & $-33.2$ [$-35.5$, $-30.9$] & $-51.5$, $-12.4$ & 1.00 & 47.3 \\
99 & 17.6 & $-23.3$ [$-24.9$, $-21.7$] & $-38.0$, $-9.4$ & 1.00 & 40.8 \\
500 & 24.9 & $-61.5$ [$-64.7$, $-58.5$] & $-87.6$, $-38.5$ & 1.00 & 86.4 \\
1234 & 16.1 & $-7.3$ [$-8.3$, $-6.3$] & $-16.4$, $+1.1$ & 0.92 & 23.4 \\
\bottomrule
\end{tabular*}
\par\smallskip{\footnotesize 2×2, TDL-A, 30 dB, 64 frames per sequence; CI: 95 percent bootstrap interval over sequences.}
\end{table}

Fixed pilots arise easily in practice. In Sionna~1.2.2
\citep{hoydis2022sionna}, the Kronecker pilot pattern draws its QPSK
pilots once, with a fixed default seed, when the resource grid is built,
so two resource grids built with default arguments carry identical pilots
(class \texttt{KroneckerPilotPattern}, argument \texttt{seed}, default 0;
a five-line reproducer is in the code repository). The neural-receiver
tutorial of the library uses this pattern, and its network, which is not
given the pilot values, is trained and evaluated with the pilots of one
such resource grid \citep{sionnatutorial}. The default 5G uplink
configuration (slot 0, cell identity 1, scrambling identity
$n_{\mathrm{SCID}}=0$) likewise
gives the same demodulation reference signal in every frame unless the
user varies the slot number or the identities. We did not check how
published studies configured these generators.

\subsection{A neural receiver of established architecture}
\label{sec:res_nrx}

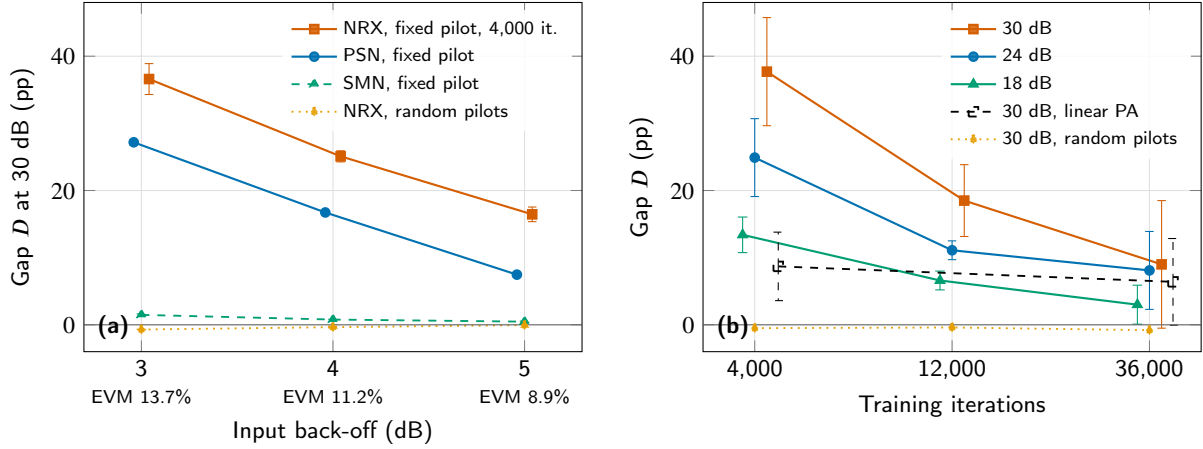
\begin{figure*}[t]
\centering
\begin{tikzpicture}
\begin{axis}[name=bo,
  width=0.47\textwidth, height=62mm,
  xmin=2.7, xmax=5.3, ymin=-4, ymax=48,
  xtick={3,4,5},
  xticklabels={{3\\[-1pt]\scriptsize EVM 13.7\%},{4\\[-1pt]\scriptsize EVM 11.2\%},{5\\[-1pt]\scriptsize EVM 8.9\%}},
  xticklabel style={align=center},
  xlabel={Input back-off (dB)}, ylabel={Gap $D$ at 30~dB (pp)},
  grid=major, grid style={black!12},
  legend style={at={(0.98,0.98)}, anchor=north east, draw=none,
                fill=white, fill opacity=0.85, text opacity=1, font=\scriptsize},
  legend cell align=left,
  tick label style={font=\small}, label style={font=\small}]
\addplot[color=cbverm, mark=square*, mark size=1.6pt, line width=0.8pt,
  error bars/.cd, y dir=both, y explicit]
  coordinates {(3.04,36.60) +- (0,2.30) (4.04,25.09) +- (0,0.83) (5.04,16.45) +- (0,1.09)};
\addlegendentry{NRX, fixed pilot, 4,000 it.}
\addplot[color=cbblue, mark=*, mark size=1.6pt, line width=0.8pt,
  error bars/.cd, y dir=both, y explicit]
  coordinates {(2.96,27.18) +- (0,0.40) (3.96,16.73) +- (0,0.24) (4.96,7.47) +- (0,0.10)};
\addlegendentry{PSN, fixed pilot}
\addplot[color=cbgreen, mark=triangle*, mark size=1.8pt, line width=0.7pt, dashed,
  error bars/.cd, y dir=both, y explicit]
  coordinates {(3.00,1.48) +- (0,0.14) (4.00,0.78) +- (0,0.04) (5.00,0.47) +- (0,0.09)};
\addlegendentry{SMN, fixed pilot}
\addplot[color=cborange, mark=diamond*, mark size=1.8pt, line width=0.7pt, dotted,
  error bars/.cd, y dir=both, y explicit]
  coordinates {(3.00,-0.70) +- (0,0.05) (4.00,-0.35) +- (0,0.16) (5.00,-0.05) +- (0,0.31)};
\addlegendentry{NRX, random pilots}
\draw[black!60] (axis cs:2.7,0) -- (axis cs:5.3,0);
\node[anchor=south west, font=\small\bfseries] at (rel axis cs:0.01,0.01) {(a)};
\end{axis}
\begin{axis}[at={(bo.south east)}, xshift=16mm, anchor=south west,
  width=0.47\textwidth, height=62mm,
  xmode=log, xmin=3000, xmax=48000, ymin=-4, ymax=48,
  xtick={4000,12000,36000}, xticklabels={{4,000},{12,000},{36,000}},
  xlabel={Training iterations}, ylabel={Gap $D$ (pp)},
  grid=major, grid style={black!12},
  legend style={at={(0.98,0.98)}, anchor=north east, draw=none,
                fill=white, fill opacity=0.85, text opacity=1, font=\scriptsize},
  legend cell align=left,
  tick label style={font=\small}, label style={font=\small}]
\addplot[color=cbverm, mark=square*, mark size=1.6pt, line width=0.8pt,
  error bars/.cd, y dir=both, y explicit]
  coordinates {(4280,37.7) +- (0,8.05) (12840,18.5) +- (0,5.35) (38520,9.0) +- (0,9.5)};
\addlegendentry{30~dB}
\addplot[color=cbblue, mark=*, mark size=1.6pt, line width=0.8pt,
  error bars/.cd, y dir=both, y explicit]
  coordinates {(4000,24.9) +- (0,5.8) (12000,11.1) +- (0,1.4) (36000,8.1) +- (0,5.8)};
\addlegendentry{24~dB}
\addplot[color=cbgreen, mark=triangle*, mark size=1.8pt, line width=0.8pt,
  error bars/.cd, y dir=both, y explicit]
  coordinates {(3740,13.4) +- (0,2.65) (11220,6.6) +- (0,1.4) (33660,3.0) +- (0,2.9)};
\addlegendentry{18~dB}
\addplot[color=black, mark=square, mark size=1.8pt, line width=0.7pt, dashed,
  error bars/.cd, y dir=both, y explicit]
  coordinates {(4560,8.7) +- (0,5.1) (41040,6.4) +- (0,6.45)};
\addlegendentry{30~dB, linear PA}
\addplot[color=cborange, mark=diamond*, mark size=1.8pt, line width=0.7pt, dotted]
  coordinates {(4000,-0.5) (12000,-0.4) (36000,-0.8)};
\addlegendentry{30~dB, random pilots}
\draw[black!60] (axis cs:3000,0) -- (axis cs:48000,0);
\node[anchor=south west, font=\small\bfseries] at (rel axis cs:0.01,0.01) {(b)};
\end{axis}
\end{tikzpicture}
\caption{Gap $D$ between the BER reduction on the training pilot and on unseen fixed pilots ($2\times2$, TDL-A).}
\label{fig:dgap}
\end{figure*}

Table~\ref{tab:nrx} and Fig.~\ref{fig:dgap}b show Check~1 for the neural
receiver. After 4,000 iterations with pilot 42, it lowers the BER of the
LMMSE baseline at 30~dB by 28.7\% on its training pilot but raises it by
7.9\% on the unseen fixed pilots and by 19.1\% with random pilots (5
seeds), so $D=36.6$~pp [34.3, 38.9], more than for the PSN after 500
iterations (27~pp) and about as much as for the PSN after 5,000
(37~pp). Trained with
random pilots for the same number of iterations, the network gains 28.2\%
on pilot 42 and 29.0\% on average on the unseen fixed pilots
($D=-0.7$~pp). A test with the training pilot
alone would thus report a gain of 29\% for the fixed-pilot receiver, where
a test with other pilots finds a loss.

\begin{table*}[t]
\caption{Neural receiver: gain on the training pilot and on unseen pilots, by training length.}
\label{tab:nrx}
\footnotesize
\begin{tabular*}{\textwidth}{@{\extracolsep{\fill}}llrrrrrrr@{}}
\toprule
Training pilots & PA & Iterations & Seeds & $R_{\mathrm{train}}$ & $R_{\mathrm{unseen}}$ & $R_{\mathrm{random}}$ & $D$ (pp) & vs.\ LIN-9 \\
\midrule
fixed (42) & Rapp & 4,000 & 5 & 28.7 & $-7.9$ & $-19.1$ & 36.6 [34.3, 38.9] & $-56.2$ \\
fixed (42) & Rapp & 12,000 & 3 & 38.6 & 20.0 & 10.9 & 18.5 [13.2, 23.9] & $-16.6$ \\
fixed (42) & Rapp & 36,000 & 3 & 48.1 & 39.1 & 30.1 & 9.0 [$-0.5$, 18.5] & 11.7 \\
\addlinespace
random & Rapp & 4,000 & 5 & 28.2 & 29.0 & 24.4 & $-0.7$ [$-0.8$, $-0.7$] & $-2.5$ \\
random & Rapp & 12,000 & 3 & 38.1 & 38.5 & 34.6 & $-0.4$ [$-0.6$, $-0.2$] & 10.5 \\
random & Rapp & 36,000 & 2 & 47.8 & 48.7 & 44.9 & $-1.5$, $-0.1$$^{a}$ & 25.8 \\
\addlinespace
fixed (42) & linear & 4,000 & 3 & 23.7 & 15.0 & 14.3 & 8.7 [3.6, 13.8] & \\
fixed (42) & linear & 36,000 & 3 & 71.3 & 64.9 & 65.4 & 6.4 [$-0.1$, 12.8] & \\
\bottomrule
\end{tabular*}
\par\smallskip{\footnotesize 2×2, TDL-A, 30 dB; R in percent relative to the LMMSE baseline; unseen = mean of pilots 123, 7 and 2024.}
\end{table*}

Longer training shrinks the gap. With three seeds per training length,
$D$ at 30~dB falls from 37.7~pp [29.7, 45.8] after 4,000 iterations to
18.5~pp [13.2, 23.9] after 12,000 and 9.0~pp [$-0.5$, 18.5] after 36,000
(Welch test between 12,000 and 36,000 iterations, $p=0.03$). At 24 and 18~dB it falls from 24.9 and 13.4~pp to 8.1~pp [2.3,
13.9] and 3.0~pp [0.1, 5.9]. The longest-trained receiver learned a
correction that also works on other pilots: on the unseen fixed pilots it
gains 39.1\% at 30~dB. On top of that it keeps a pilot-specific part.
After 36,000 iterations $D$ was positive in all nine combinations of seed
and SNR, between 2.0 and 13.3~pp, although with three seeds the 95\%
interval at 30~dB includes zero. The validation loss on the held-out
pilot fell from 0.093 after 4,000 iterations to 0.066 after 12,000 and
0.057 after 36,000, and it changed by less than 1\% over the last fifth
of the longest run, while the loss on the training pilot stayed lower
(0.043). The learning rate is small at the end of a cosine schedule,
however, so this flattening does not prove convergence, and since $D$
was still falling between 12,000 and 36,000 iterations, even longer
training might reduce it further.

Training on one pilot also costs accuracy on others. Trained with random
pilots for 36,000 iterations, the receiver gains 48.7\% on the unseen fixed pilots at 30~dB, against 39.1\%
after fixed-pilot training, a difference of 9.6~pp at 30~dB, and of 6.6
and 3.5~pp at 24 and 18~dB. The random-pilot
receiver itself has no gap ($-1.5$ and $-0.1$~pp for its two seeds at
30~dB).

Part of the gap does not come from the PA. With a linear PA, the
fixed-pilot receiver had $D=8.7$~pp [3.6, 13.8] after 4,000 iterations and
6.4~pp [$-0.1$, 12.8] after 36,000 (3 seeds each). Subtracting this seed
by seed, the part of the gap attributable to the PA fell from 28.8~pp
after 4,000 iterations (for the three seeds of the first experiment that
were also run with a linear PA, whose $D$ with the Rapp PA is 37.5~pp) to
2.7~pp [$-5.1$, 10.4] after
36,000 at 30~dB; at 18 and 24~dB it was 1.4~pp [0.3, 2.5] and 5.1~pp
[1.3, 9.0]. After long training, most of the remaining gap at 30~dB
therefore does not come from the PA. We did not
identify the pilot-specific structure that the network uses without a
PA. One candidate is inter-carrier interference from phase noise, which
adds to the LS estimate on subcarrier $k$ terms proportional to
$q_m/q_k$ for neighbouring subcarriers $m$. The interference coefficients
are random, but for a fixed pilot these ratios are the same in every
frame. The
PSN shows no such effect (Section~\ref{sec:res_controls}), possibly
because it acts after the equaliser and the tracker with one
multiplicative correction per subcarrier.

How does the neural receiver compare with the training-free estimates?
After 4,000 iterations with random pilots it was 1.7 to 4.4\% worse than
LIN-9 smoothing at 30~dB on every pilot and 43 to 49\% worse at 18~dB.
After 36,000 iterations its BER on unseen pilots was 25.8\% below that of
LIN-9 at 30~dB and 34\% below that of the PA-aware estimate; it equalled
LIN-9 at 24~dB ($+1.7\%$) and was 32.6\% worse at 18~dB. Trained with the
fixed pilot for 36,000 iterations, it beat LIN-9 at 30~dB by 24.8\% on
its training pilot but by only 11.7\% [$-2.1$, 25.5] on unseen pilots.

\subsection{The gap at lower amplifier distortion}\label{sec:res_backoff}

Fig.~\ref{fig:dgap}a shows $D$ at 30~dB as the PA is backed off (5 seeds
each). With a data EVM of 13.7, 11.2 and 8.9\% at IBO 3, 4 and 5~dB, the
gap of the PSN falls from 27.2 to 16.7 and 7.5~pp, and that of the neural
receiver after 4,000 iterations from 36.6 to 25.1 and 16.4~pp, while the
SMN and the receivers trained with random pilots stay at or below 1.5~pp.
At 4~dB the EVM meets the 12.5\% that 3GPP allows a base station for
16-QAM \citep{3gpp38104}, so the effect is not confined to an EVM outside
the standard. We did not evaluate the adjacent-channel leakage at 4~dB; at
3~dB it is about 20~dB above the base-station limit
(Section~\ref{sec:res_pilot}). The neural-receiver values at 4 and 5~dB
are short-training values. At 3~dB, longer training reduced the gap about
fourfold, and we did not repeat the long training at 4 and 5~dB.

\subsection{Controls without pilot distortion}\label{sec:res_controls}

Table~\ref{tab:controls} gives Check~2 for the stages. With PA-distorted
pilots the PSN beats the SMN by 20.7~pp [20.0, 21.3] at 30~dB. With the
true channel or with clean pilots the advantage disappears ($-0.1$~pp),
because there is no $b_k$ pattern to store; per seed it stays within
0.4~pp of zero on both channels and at all three SNRs, against means of
9.5 to 20.7~pp with the estimated channel. With clean pilots the estimate lacks
the gain $\alpha$ of the data, so both stages recover 28\% with an
essentially global correction; scaling the clean-pilot estimate by
$\alpha$ lowers the baseline BER from 0.036 to 0.020 (768 frames), close
to the 0.018 of the PA-aware estimate.

\begin{table}[t]
\caption{Check 2: BER reduction of the PSN and the SMN without pilot distortion.}
\label{tab:controls}
\footnotesize
\setlength{\tabcolsep}{3pt}
\begin{tabular*}{\columnwidth}{@{\extracolsep{\fill}}lrrrr@{}}
\toprule
Condition & BER$_0$ & PSN & SMN & PSN $-$ SMN \\
\midrule
Estimated channel & 0.0239 & 22.1$\pm$0.7 & 1.4$\pm$0.1 & 20.7$\pm$0.7 \\
Ideal channel & 0.1050 & 7.1$\pm$0.4 & 7.1$\pm$0.4 & $-$0.06$\pm$0.06 \\
Clean pilots & 0.0359 & 28.2$\pm$0.5 & 28.4$\pm$0.6 & $-$0.14$\pm$0.08 \\
\bottomrule
\end{tabular*}
\end{table}

The ideal-channel baseline has a high BER (0.105) because of phase
tracking, not because of missing gain: scaling by $1/|\alpha|$ gives
0.086 and the genie channel $\alpha\mathbf{H}_k$ 0.085, whereas without
phase noise the true-channel receiver reaches 0.0075. A pilot-based
estimate captures each oscillator's phase at the pilot symbols and leaves
the tracker a small residual; with the true channel the tracker must
follow the whole Wiener drift (a standard deviation of about 0.6~rad over
a frame). A sequential tracker, which starts each symbol from the phase of
the previous one, improves the genie receiver from 0.084 to 0.026 (512
frames, a separate run), still above the estimated channel (0.019). It
also improves the estimated-channel receivers (plain 0.023 to 0.019,
LIN-9 0.017 to 0.013, PA-aware 0.018 to 0.015) without changing their
order. The gain, SMN and PSN all reach 7\% in the ideal-channel pipeline,
the same global correction, so we do not count this condition as
evidence; the clean-pilot condition is the informative control. For the
neural receiver, the corresponding control is the linear PA of
Section~\ref{sec:res_nrx}.

\subsection{Training-free estimates and what is left for learning}
\label{sec:res_smooth}

On the PSN's training sequence the PA-aware estimate matches the network
at $2\times2$ ($+0.1$~pp [$-0.3$, $+0.5$]) and is slightly better at
$4\times4$ ($+0.5$~pp [$+0.1$, $+0.9$]). Unlike the network, it lowers
the BER at 30~dB on every sequence tested: by 15 to 27\% in the runs with
trained receivers, to an absolute BER of 0.018 to 0.019 at $2\times2$
whichever sequence is sent (Fig.~\ref{fig:pilot_test}), and by at least
10.5\% on all 107 sequences of Section~\ref{sec:res_psn}. Because it removes $\mu_k$ and nothing else,
a network that matches it on the training sequence and fails on all
others has learned the pilot-specific ratio $b_k/\alpha$, not a
compensation of the PA. Trained on top of the PA-aware estimate, the
stages add at most 0.3~pp at $\mathrm{IBO}=3$~dB: at $4\times4$ the
8-parameter gain adds $+0.22$~pp [$+0.13$, $+0.31$], the SMN $+0.05$~pp
and the cubic term nothing beyond its interval, and the 27,650-parameter
PSN lowers the reduction by 0.34~pp [0.26, 0.42]. Across both PAs and all
back-offs, no stage removes more than 1.2\% of the errors left by the
PA-aware estimate (Fig.~\ref{fig:pa_backoff}b), and the SMN never beats the
4-parameter gain. The remaining in-band distortion $d_k$ arises from
clipping the time-domain sum of all subcarriers and is nearly unrelated to
$x_k$ itself \citep{dardari2000}, so a stage that acts on one resource
element or a small neighbourhood cannot predict it; removing it requires
processing the whole OFDM symbol \citep{chen2003}. Our results are
consistent with this explanation but do not prove it. The neural
receiver does improve on the PA-aware estimate after long training
(Section~\ref{sec:res_nrx}), but we did not determine whether it reduces
the distortion or copes better with noise and phase noise.

A standard step that needs no PA model does better than the PA-aware
estimate (Table~\ref{tab:smooth}). Smoothing the plain LS estimate with
LIN-9 or LM-100 lowers the BER of the LMMSE baseline at 30~dB by 30 to
36\% ($2\times2$) and 27 to 31\% ($4\times4$) on every test sequence,
more than the PSN achieves on its training sequence (22 and 14.5\%) and
more than the PA-aware estimate in the same runs (15 to 26\%). In the
same study, on the i.i.d.\ and TDL-C (100~ns) channels the reductions are
24 to 34\% ($2\times2$) and 18 to 34\% ($4\times4$). Smoothing the PA-aware estimate gives nearly the same
BER as smoothing the plain one (within 1~pp for fixed sequences, up to
5~pp for random pilots). The default TDL-A channel is nearly flat over the
band (correlation 0.5 out to about 300 subcarriers), which favours
smoothing.

\begin{table*}[t]
\caption{Smoothed and PA-aware estimates, with and without learned stages on top.}
\label{tab:smooth}
\footnotesize
\begin{tabular*}{\textwidth}{@{\extracolsep{\fill}}lrrrr@{}}
\toprule
Test pilot sequence & 42 (training) & 123 & 2024 & random \\
\midrule
\multicolumn{5}{@{}l}{\textit{$2\times2$}}\\
Plain LMMSE BER ($\times10^{-2}$) & 2.40 & 2.54 & 2.43 & 2.51\\
PSN on plain LS & 22.0$\pm$0.7 & $-$2.5$\pm$0.3 & $-$9.7$\pm$0.4 & $-$18.4$\pm$6.3\\
LIN-9 smoothing, no training & 30.2$\pm$2.9 & 34.7$\pm$0.8 & 31.4$\pm$4.4 & 31.4$\pm$1.8\\
LM-100 smoothing, no training & 30.9$\pm$2.9 & 35.7$\pm$2.4 & 32.5$\pm$4.3 & 33.0$\pm$1.2\\
PA-aware LS, no training & 21.5$\pm$0.2 & 26.1$\pm$1.1 & 22.7$\pm$2.7 & 25.9$\pm$5.3\\
PSN on LIN-9 & 31.3$\pm$3.0 & 33.7$\pm$0.9 & 30.7$\pm$4.3 & 29.6$\pm$2.3\\
PSN on LM-100 & 31.0$\pm$2.7 & 34.2$\pm$2.5 & 32.5$\pm$4.2 & 32.9$\pm$0.9\\
SMN on LIN-9 & 30.1$\pm$2.7 & 34.4$\pm$0.9 & 31.1$\pm$4.4 & 31.4$\pm$1.4\\
\addlinespace
\multicolumn{5}{@{}l}{\textit{$4\times4$}}\\
Plain LMMSE BER ($\times10^{-2}$) & 3.57 & 3.75 & 3.63 & 3.69\\
PSN on plain LS & 14.5$\pm$1.2 & $-$2.5$\pm$0.5 & $-$6.8$\pm$0.4 & $-$13.6$\pm$3.2\\
LIN-9 smoothing, no training & 26.5$\pm$2.3 & 30.1$\pm$1.8 & 28.0$\pm$1.8 & 28.5$\pm$6.3\\
LM-100 smoothing, no training & 27.3$\pm$2.9 & 31.4$\pm$1.5 & 29.0$\pm$2.1 & 30.0$\pm$6.5\\
PA-aware LS, no training & 15.0$\pm$0.7 & 16.5$\pm$1.6 & 14.7$\pm$1.4 & 16.9$\pm$6.2\\
PSN on LIN-9 & 27.2$\pm$2.3 & 29.4$\pm$1.6 & 27.6$\pm$1.8 & 27.4$\pm$6.2\\
PSN on LM-100 & 27.3$\pm$3.0 & 30.4$\pm$1.4 & 29.0$\pm$2.1 & 29.7$\pm$6.5\\
SMN on LIN-9 & 26.6$\pm$2.3 & 30.0$\pm$1.7 & 28.0$\pm$1.9 & 28.6$\pm$6.2\\
\bottomrule
\end{tabular*}
\end{table*}

Table~\ref{tab:budget} shows what smoothing removes. The link is limited
by phase noise, which triples the BER. With the PA and all impairments,
smoothing gains 28.5\%, and the PA-aware estimate, which removes the
ripple exactly and nothing else, 20.6\%. Smoothing the PA-aware estimate
adds 11\%, which is thermal-noise averaging: it falls to 2.4\% without
thermal noise and vanishes without thermal and phase noise, where the
PA-aware estimate is exact (normalised error $-85$~dB). With a linear PA
and phase noise the same smoothing loses 3.8\%, and without phase noise
it gains 56\%, ordinary noise averaging; with clean pilots it gains 8 to
15\%. The effects are therefore not additive. In this phase-noise-limited
link most of the smoothing gain is removal of the pilot ripple, and the
rest is noise averaging that shows only with the PA. The result does not
depend on the smoother: in a separate 512-frame run, all windows from 5 to
33 subcarriers give 22 to 31\% on TDL-A at 30~dB, and LM-100 gives 25 to
31\% for $\sigma_s^2$ from 0.001 to 0.05.

\begin{table}[t]
\caption{Error budget: smoothing and PA-aware gains with impairments switched off.}
\label{tab:budget}
\footnotesize
\setlength{\tabcolsep}{2pt}
\begin{tabular*}{\columnwidth}{@{\extracolsep{\fill}}lrrrr@{}}
\toprule
Condition & Plain BER & LIN-9 & PA-aware & LIN-9 on PA-aw. \\
\midrule
All impairments & 2.47 & 28.5 & 20.6 & 11.0\\
No PN & 0.76 & 70.3 & 54.3 & 40.3\\
No thermal noise & 2.14 & 22.5 & 21.4 & 2.4\\
No PN, no thermal & 0.42 & 78.4 & 82.2 & $-$0.1\\
Linear PA & 1.75 & $-$3.8 & 0.0 & $-$3.8\\
Linear PA, no PN & 0.11 & 56.2 & 0.0 & 56.2\\
Clean pilots & 3.75 & 8.3 & -- & --\\
Clean pilots, no PN & 0.98 & 15.3 & -- & --\\
\bottomrule
\end{tabular*}
\end{table}

With the ripple removed, little is left to memorise. Trained with
sequence 42 on LIN-9 or LM-100, the PSN gains at most 1.6\% over its own
smoothed baseline on the training sequence and loses up to 2.6\% on unseen
ones. Its gap (unseen pilots 123 and 2024, 3 and 4 seeds) falls from
28.1~pp [27.5, 28.8] on the plain estimate to
2.8~pp [2.6, 2.9] on LIN-9 and 1.3~pp [1.2, 1.5] on LM-100 at $2\times2$,
and from 19.2~pp [18.0, 20.4] to 1.8~pp [1.5, 2.1] and 0.7~pp [0.3, 1.1]
at $4\times4$; for the SMN it is 0.3 to 0.4~pp. With 5,000 instead of 500
iterations the picture does not change: on LIN-9 the gap is 2.9~pp for the
PSN and 0.4~pp for the SMN at 30~dB, and neither gains more than 2.0\% on
the training pilot (3 seeds). On the plain estimate, longer training lets
the SMN gain more than before (6.9\% on the training pilot and 5.0\% on
unseen pilots at 30~dB, against about 1.4\% after 500 iterations), still
far below smoothing. Coded transmission with the low-density parity-check
(LDPC) code of 5G (256 codewords per value) confirms the order at 22~dB
and rate 3/4: the block error rate is 0.22 to
0.24 with the plain estimate, 0.14 to 0.17 with LIN-9 and 0.20 to 0.21
with the PA-aware estimate. The memoriser of pilot 42 equals the PA-aware
estimate on its training pilot (0.21) and is worst on unseen pilots
(0.29). At rate 1/2 all receivers stay at or below 3\% from 22~dB.

\subsection{Delay spread: where smoothing stops working}
\label{sec:res_delay}

With the correlation $\Sigma_{kk'}$, subcarriers $\Delta k$ apart are
correlated by 0.5 at about 92, 31 and 9 subcarriers for 100, 300 and
1000~ns, and by 0.9 at about 26, 9 and 3, against a nine-subcarrier LIN-9
window. Fixed windows therefore fail from 300~ns
(Table~\ref{tab:delay}): LIN-9 increases the BER by 27\% on TDL-C at
300~ns and by a factor of five or more at 1000~ns, and LM-100 collapses
at 300~ns. A smoother matched to the true $\tau$ degrades more gracefully
but loses most of its benefit at 30~dB: 19, 5 and $-14\%$ at 300, 600 and
1000~ns on TDL-C, and 24 and 12\% at 300 and 1000~ns on TDL-A. The
PA-aware estimate, which does not average across subcarriers, gains 22
to 23\% at every delay spread. Where smoothing fails, the pilot
fingerprint returns. On top of the matched smoother on TDL-C, the PSN
trained on sequence 42 has $D=1.9$~pp [1.6, 2.3] at 300~ns but 5.0~pp
[4.8, 5.1] at 1000~ns (5 seeds each), with a negative unseen-pilot result
in all five seeds. In the same 5-seed runs at 1000~ns and
30~dB, neither the matched smoother ($-12.4\%$) nor the PSN on it
($-14.5\%$) beats the plain baseline on the unseen pilots 123, 2024 and
random. Smoothing is an adequate reference only for
short delay spreads, and Check~1 remains necessary. At 1000~ns the last
TDL taps exceed the 8.33~$\mu$s cyclic prefix, so these cases test
estimation only.

\begin{table}[t]
\caption{Training-free estimates versus delay spread.}
\label{tab:delay}
\footnotesize
\setlength{\tabcolsep}{2pt}
\begin{tabular*}{\columnwidth}{@{\extracolsep{\fill}}lrrrrrr@{}}
\toprule
Channel & $\tau$ (ns) & BER$_0$ (\%) & LIN-9 & LM-100 & LM-$\tau$ & PA-aware \\
\midrule
TDL-C & 100 & 2.41 & 26 & 23 & 23 & 22\\
TDL-C & 300 & 2.50 & $-$27 & $-$305 & 19 & 23\\
TDL-C & 600 & 2.48 & $-$285 & $-$507 & 5 & 22\\
TDL-C & 1000 & 2.41 & $-$491 & $-$765 & $-$14 & 22\\
TDL-A & 300 & 2.55 & $-$11 & $-$337 & 24 & 23\\
TDL-A & 1000 & 2.36 & $-$646 & $-$947 & 12 & 23\\
\bottomrule
\end{tabular*}
\end{table}

\subsection{Pilot design, PA sampling and back-off}
\label{sec:res_pilot}

Table~\ref{tab:pilot} shows how the pilot error depends on the pilot
design and on how the PA is simulated. With a $4\times$ oversampled PA
and a guard band, the ripple of random BPSK pilots falls from 0.138 to
0.110: critical sampling overstates the ripple by about a quarter but does
not create it. In every case the ripple equals the data EVM within 10\%
(Rapp 1 to 9~dB, Saleh 3 to 9~dB and an ideal soft limiter, each at
$1\times$, $4\times$ and $4\times$ with guard band), the mismatch between
two sequences is close to $\sqrt2$ times the EVM for the Rapp PA (21.5
against 19.4\%), and after LIN-9 the ripple falls by the factor predicted
for ripple that is independent between subcarriers (to 4.9\%; predicted
5.1\%). The size of the pilot error can thus be predicted from the
transmitter EVM without simulating the receiver.

\begin{table*}[t]
\caption{Pilot ripple and Check 1 for BPSK, QPSK and Zadoff–Chu pilots.}
\label{tab:pilot}
\footnotesize
\begin{tabular*}{\textwidth}{@{\extracolsep{\fill}}lrrrrrrrrr@{}}
\toprule
Pilot family & \multicolumn{2}{c}{Ripple of $b_k/\alpha$} & Plain BER & LIN-9 & PA-aware & \multicolumn{3}{c}{PSN, fixed-pilot training} & $D$ (pp) \\
\cmidrule(lr){2-3}\cmidrule(lr){7-9}
 & $1\times$ & $4\times$, guard & ($\times10^{-2}$) & & & training & unseen fixed & random & \\
\midrule
BPSK $\pm1$ & 0.138 & 0.110 & 2.41 & 30.8 & 22.7 & 21.9 & $-$5.0 & $-$15.0 & 26.8 [26.5, 27.1]\\
QPSK & 0.137 & 0.109 & 2.83 & 43.7 & 36.8 & 38.8 & $-$32.7 & $-$35.7 & 71.5 [69.8, 73.2]\\
Zadoff--Chu & 0.000 & 0.040 & 2.55 & 7.6 & 30.1 & 15.3 & 15.2 & 15.0 & 0.1 [$-$1.5, 1.7]\\
\bottomrule
\end{tabular*}
\end{table*}

QPSK pilots, like the QPSK of the NR reference signal \citep{3gpp38211},
have the same complex ripple as BPSK pilots but with a phase part (real
and imaginary parts of the relative deviation 9.5 and 10.0\%, against
14.0 and 1.7\% for BPSK). Trained on one QPSK pilot, the PSN gains 38.8\%
on it and loses 31 to 36\% on unseen QPSK pilots, $D=71.5$~pp, almost
three times the BPSK value, while the SMN shows no gap ($D=-0.2$~pp). The
memoriser separates amplitude and phase (20 unseen pilots per family):
for QPSK, storing only the phase gives $D=55$~pp and only the amplitude
30~pp, whereas for BPSK the amplitude gives 55 and the phase 2~pp. A
per-subcarrier phase error therefore drives the larger QPSK loss,
plausibly because the common-phase tracker cannot remove it. Constant-envelope
Zadoff--Chu pilots \citep{chu1972} of length 64 occupy all 64
subcarriers, so at critical sampling their time signal receives the same
complex gain on every sample. Their ripple vanishes (0.02 to 0.04 with
oversampling), and with it, within the resolution of these runs, the gap
(memoriser: $D=-2$~pp; trained PSN: $D=0.1$~pp [$-1.5$, $1.7$]). What
remains is a common gain offset ($|\bar b|/|\alpha|=1.11$, with $\bar b$
the mean of $b_k$), which the PA-aware estimate removes, the stages
correct only in part (15 against 30\% in Table~\ref{tab:pilot}) and
smoothing cannot correct. NR reference signals mapped to a comb are not
constant-envelope inside a larger FFT. In an NR-like configuration
(1,024-point FFT with 624 occupied subcarriers, QPSK; NumPy chain), a
whole-symbol pilot has a deterministic ripple of 12.1\%, as large as the
data EVM there (11.5\%); a comb-2 reference signal with data on the other
resource elements keeps a deterministic ripple of 5.7\%, plus 10.0\% that
changes with the data, and a reference-signal symbol without data and
with the 3~dB power boost of NR keeps 11.4\%. Once data share the pilot
symbol, the PA-aware estimate can no longer be computed.

\begin{figure*}[t]
\centering
\includegraphics[width=\textwidth]{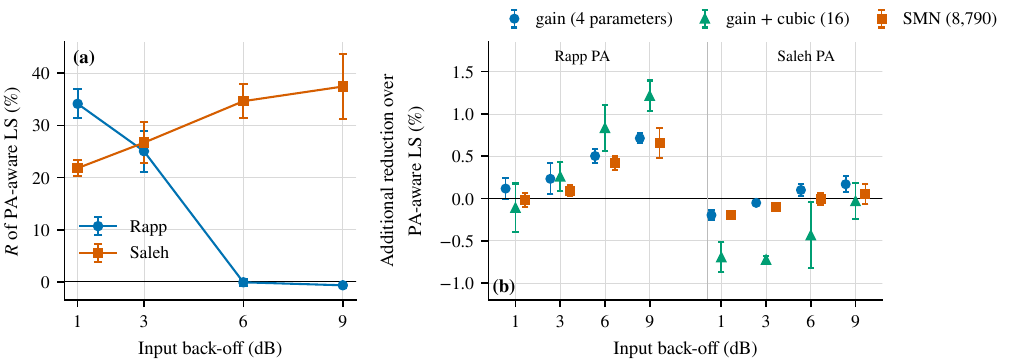}
\caption{ (a) BER reduction of PA-aware LS relative to plain LS; (b) additional reduction of learned stages trained on top of it }
\label{fig:pa_backoff}
\end{figure*}

The default operating point is aggressive. At 3~dB IBO the Rapp PA gives
a data EVM of 13.7\% at critical sampling, slightly above the 12.5\% that
3GPP allows a base station for 16-QAM \citep{3gpp38104}, and 11.2\% with
oversampling; at 3~dB the adjacent-channel leakage ratio of the
oversampled chain is about $-25$~dBc, about 20~dB worse than the 45~dB
required of a base station. Fig.~\ref{fig:pa_backoff}a shows the gain of
the PA-aware estimate over the plain one versus back-off: 34 and 25\% for
the Rapp PA at 1 and 3~dB but zero at 6 and 9~dB, and 22 to 37\% for the
Saleh PA at every back-off. The data EVM is 19.4, 13.7, 6.8 and 2.7\% for
the Rapp PA at 1, 3, 6 and 9~dB, but 32.9, 22.9 and 15.7\% for the Saleh
PA at 3, 6 and 9~dB; the Saleh model is thus a strongly compressing
stress case, and the two PAs are not compared at equal operating points.
The memoriser shows where the gap ends (pilot 42, 20 unseen BPSK pilots,
30~dB): $D$ is 59~pp at 3~dB (55~pp over all 106, Table~\ref{tab:many}),
31~pp at 4~dB, where every unseen pilot loses, 13~pp at 5~dB, with 95\%
of unseen pilots losing, and $-1$~pp at 6~dB. For the Saleh PA it is 60
and 77~pp at 6 and 9~dB. These training-free values agree in trend with
the trained receivers of Fig.~\ref{fig:dgap}a. Our results describe a PA
without predistortion in or near strong compression; a linearised
transmitter at a comfortable back-off would show a small pilot error and
little gap.

\subsection{Limits of the PA-aware estimate}\label{sec:res_limits}

\begin{figure*}[t]
\centering
\includegraphics[width=\textwidth]{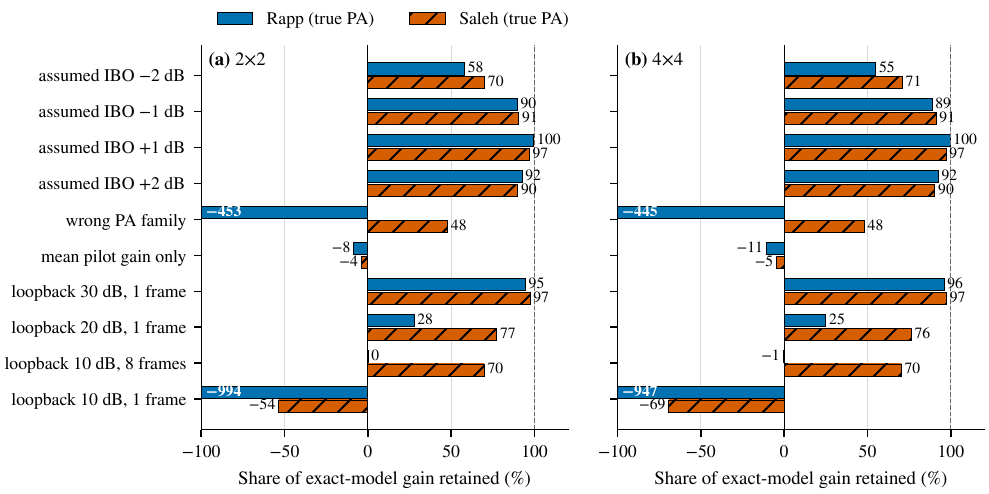}
\caption{Share of the PA-aware gain retained with a wrong PA model or a loopback reference .}
\label{fig:robust}
\end{figure*}

The PA-aware estimate removes the ripple exactly only because it is given
the true PA model. The simulated numbers indicate trends, not hardware
tolerances. We express each variant as the retained share
$\eta=100\,R_{\mathrm{variant}}/R_{\mathrm{exact}}$~\%, where
$R_{\mathrm{exact}}$ is the reduction with the true model, averaged over
sequences 42 and 123 at 30~dB: 24.5\% (Rapp) and 27.1\% (Saleh) at
$2\times2$ and 16.5 and 23.2\% at $4\times4$ (Fig.~\ref{fig:robust}; both
array sizes agree closely). A back-off error of up to 1~dB retains 89 to
100\% of the gain, 2~dB too little back-off 55 to 71\% and 2~dB too much
90 to 92\%; errors in the Rapp smoothness ($\nu=1$ or 4) or ignoring AM/PM
retain 93 to 99\%. A wrong PA family is critical: a Saleh reference for a
Rapp PA doubles the baseline BER ($\eta\approx-450\%$), because it
predicts a ripple more than twice as large with strong phase rotation,
while a Rapp reference for a Saleh PA retains 48\%. Correcting only the
mean pilot gain gives nothing ($-4$ to $-11\%$). Instead of a model,
$\tilde{\mathbf{q}}$ and $\alpha$ can be measured by a loopback, for
example with the observation receiver of a predistortion system
\citep{morgan2006}. At a calibration SNR of 30~dB (one frame) 95 to 97\%
of the gain is retained, at 20~dB only 25 to 28\% (Rapp) and 76 to 77\%
(Saleh), because calibration noise enters every later estimate as the same
fixed error and must be well below the ripple. Re-estimating the channel
from the hard decisions of a first pass, without PA knowledge, increased
the BER by 12 to 16\% at 30~dB. The transmitter could also remove the
ripple at the source, by predistorting the pilot symbols alone or by
sending them with more back-off; we did not simulate these options.

\section{Discussion}\label{sec:discussion}

\subsection{Why a fixed pilot overstates the gain}

When pilots pass through a nonlinear PA and one sequence is used for
training and testing, the channel estimate carries a fixed,
sequence-specific error that a receiver with enough freedom can learn to
undo; only constant-envelope pilots at critical sampling avoid it. The resulting gain is a property of the test setup, not of the
receiver. The error is large only where the PA distorts the pilots
strongly, at low back-off or for gradually compressing PAs with strong
AM/PM, which is where learned receivers are proposed. Because the ripple
matches the data EVM, the transmitter EVM tells in advance how large the
pilot error is. It does not rule out a gap, however: the neural receiver
showed a smaller one even with a linear PA (Section~\ref{sec:res_nrx}).

Both learned receivers learned the pattern. For the PSN this was
expected: it has its own matrices for every subcarrier, and its gap grew
with training. The neural receiver shares its convolution weights across the
grid, yet after short training its gap was as large as that of the PSN
after long training.
It has two ways to locate a subcarrier: its layer normalisations carry a
learned scale and offset for every resource element, and the band edges
of the zero-padded grid mark the positions near them. Its input also contains the LS
estimate, which carries the fingerprint itself. We did not test which of
these it uses; an ablation with normalisation parameters shared across
the grid would show whether the per-element parameters matter. With long
training its gap fell to 3 to 9~pp, and at 30~dB most of what remained
also appeared with a linear PA; the part attributable to the PA had a
95\% interval above zero only at 18 and 24~dB (1.4 and 5.1~pp). For a well-trained
receiver of this type the overstatement is therefore much smaller than
after short training, and only partly caused by the PA, but it did not
vanish within 36,000 iterations, and training on one pilot still cost
accuracy on others.

\subsection{Recommended evaluation protocol}

For any learned receiver evaluated under PA distortion we recommend the
following.
\begin{enumerate}
\item Hold out what varies in deployment, here the pilot sequence (in NR
the reference signal changes with the slot and the cell
\citep{3gpp38211}), train with changing pilots, and report $D$ with its
spread over sequences.
\item Report a control without pilot distortion: pilots that bypass the
PA, scaled by $\alpha$, or a linear PA. An ideal-channel control can be
confounded by phase tracking.
\item Compare against a smoothed LS estimate with a window suited to the
delay spread, a PA-aware estimate where the PA is known, and minimal
models such as a single complex gain, and give the absolute BER of every
reference.
\item Report the training length and the validation loss on a held-out
pilot, since the gap of the neural receiver changed fourfold with training
length.
\item State how the pilots were generated: family, generator and seed,
and whether they are fixed or changing.
\end{enumerate}
As a pass criterion we suggest that $D$ stays within a stated margin, for
example 1~pp, the seed-to-seed spread of our stage experiments, tested as
an equivalence and not as a difference from zero. The margin depends on
the baseline and the SNR, and $D\approx0$ is necessary but not
sufficient. Applied to our own data, the SMN on LIN-9 passes ($D=0.3$ to
0.4~pp) and the SMN on plain LS narrowly fails ($D=1.5$~pp). The PSN on
LIN-9 fails ($D=2.8$ and 1.8~pp at $2\times2$ and $4\times4$), and so
does the neural receiver after 36,000 iterations ($D=3.0$ to 9.0~pp),
while the Zadoff--Chu result is inconclusive (0.1~pp [$-1.5$, 1.7]).

The choice of baseline matters as well. Against plain LS, the PSN
appears to gain 14 to 22\%; against a smoothed estimate, none of the
stages we placed after the tracker gained more than 2\% on its training
pilot, even with ten times longer training. The neural receiver is the exception: after
long training with changing pilots it beat smoothing by 26\% at 30~dB,
matched it at 24~dB and lost to it at 18~dB. A comparison with plain LS
alone would have hidden this dependence on SNR, and a test with the
training pilot alone would have hidden the pilot effect.

\subsection{Applicability}\label{sec:disc_appl}

On hardware, Check~1 needs only a reconfigurable pilot, such as the slot
number or scrambling identity of an NR reference signal or the pilot seed
of a software radio. The PA-aware part of Check~3 needs a PA model or a
loopback measurement. The clean-pilot control exists only in simulation,
but pilots sent with extra back-off approximate it. The convolutional
receivers DeepRx and HybridDeepRx \citep{honkala2021deeprx,pihlajasalo2023}
also receive the LS estimate, which carries the pattern; whether they
exploit it depends on their architecture and training, and the checks
apply to them unchanged. A neural receiver that, like the Sionna
tutorial network, is not given the pilot values can only work with the
pilots it was trained on, so changing pilots must be part of its design
before Check~1 can be applied. Some systems do use fixed training
sequences, such as the long training field of Wi-Fi. There the pattern
also exists in deployment, but it depends on the PA, its back-off and its
AM/PM (Sections~\ref{sec:pilotpa} and~\ref{sec:res_pilot}), which differ
between transmitters and are unknown to the receiver. A receiver that
learns the pattern from one simulated PA learns a correction for that
PA, so it should be tested across PA models and back-offs.

\appendix
\setcounter{table}{0}
\renewcommand{\thetable}{A.\arabic{table}}
\renewcommand{\theHtable}{A.\arabic{table}}
\section{List of experiments}\label{app:exps}

Table~\ref{tab:exps} lists the experiments, their configurations and the
numbers of seeds. Unless stated otherwise, they use the Rapp PA at
$\mathrm{IBO}=3$~dB.

\begin{table*}[!t]
\caption{Experiments and numbers of seeds or replicates (Rapp PA at 3~dB unless stated). NRX: neural receiver.}
\label{tab:exps}
\footnotesize
\begin{tabular*}{\textwidth}{@{\extracolsep{\fill}}Y{0.19\textwidth}Y{0.34\textwidth}Y{0.31\textwidth}Y{0.08\textwidth}@{}}
\toprule
Experiment & Configuration & Methods compared & Seeds \\
\midrule
Pilot test (Check 1) & $2\times2$, TDL-A and i.i.d.; test pilots 42, 123, 7, 2024, random & PSN and SMN, fixed-pilot and random-pilot training & 10 \\
Block sharing & $2\times2$, TDL-A; pilot 42 & PSN with matrices shared in blocks of 1 to 64 subcarriers & 20 \\
Stage training length & $2\times2$, TDL-A; 500 and 5,000 iterations; plain LS and LIN-9 inputs & PSN, SMN (fixed pilot) & 3 \\
Neural receiver & $2\times2$, TDL-A; 4,000 iterations; Rapp PA and linear PA & NRX, fixed-pilot and random-pilot training & 5 (linear: 3) \\
NRX training length & $2\times2$, TDL-A; 4,000, 12,000, 36,000 iterations; linear PA at 36,000 & NRX, fixed-pilot and random-pilot training & 3 (random, 36,000: 2) \\
Trained back-off & $2\times2$, TDL-A; IBO 3, 4, 5~dB & PSN, SMN (500 it.), NRX (4,000 it.), fixed and random pilots & 5 \\
Controls (Check 2) & $2\times2$, TDL-A and i.i.d.; ideal channel, clean pilots & PSN, SMN; diagnostics with the true channel, $\alpha\mathbf{H}_k$, sequential tracker (768 or 512 frames) & 5 \\
PA-aware comparison, $2\times2$ (Check 3) & TDL-A and i.i.d.; pilots 42, 123, 2024, random & PSN (fixed pilot); PSN, SMN, gain, cubic (random pilots) on PA-aware LS; decision-directed re-estimation & 5 \\
PA-aware comparison, $4\times4$ (Checks 1, 3) & TDL-A; pilots 42, 123, random & PSN (fixed pilot); PA-aware LS alone and with gain, cubic, SMN or PSN & 6 \\
PA and back-off & $2\times2$, TDL-A, Rapp and Saleh at 1, 3, 6, 9~dB; random test pilots & PA-aware LS alone and with gain, cubic or SMN & 4 \\
Sensitivity of PA-aware LS & TDL-A, Rapp and Saleh 3~dB; pilots 42, 123; 768 ($2\times2$) and 192 ($4\times4$) frames per value & wrong PA assumptions, mean-only correction, loopback calibration & 6, 5 \\
Smoothing (Checks 1, 3) & $2\times2$ and $4\times4$, TDL-A, TDL-C (100~ns), i.i.d.; pilots 42, 123, 2024, random & LIN-$M$, moving averages, LM-100/300, PA-aware LS with and without LIN-9; PSN and SMN on LIN-9 and LM-100 & 3, 4 \\
Delay spread & $2\times2$, TDL-C at 100 to 1000~ns, TDL-A at 300, 1000~ns; pilots 42, 123, 2024, random & smoothers incl.\ LM-$\tau$ matched to the delay spread, PA-aware LS; PSN on the matched smoother and on LIN-9, SMN on LIN-9 & 5 (training-free sweep: 1) \\
Pilot design & $2\times2$, TDL-A, TDL-C (100~ns), i.i.d.; BPSK, QPSK, Zadoff--Chu; linear PA, clean pilots; 512 frames & plain, smoothed (window and $\sigma_s^2$ sweeps), PA-aware LS; PSN and SMN trained on one fixed QPSK or Zadoff--Chu pilot & 3 \\
Pilot ripple (NumPy) & Rapp 1--9~dB, Saleh 3--9~dB, soft limiter; $1\times$, $4\times$, $4\times$ with guard band; NR-like comb reference signals & ripple of $b_k/\alpha$ against data EVM & 200 sequences \\
Error budget, trackers & $2\times2$, TDL-A, 30~dB; impairments switched off in turn; 512 frames & plain, LIN-9, PA-aware LS; original and sequential tracker & 1 \\
Memoriser (Check 1) & $2\times2$, TDL-A, 30~dB (also 24~dB); Rapp 3--6~dB, Saleh 6, 9~dB; BPSK, QPSK, Zadoff--Chu; 32 or 64 frames per sequence (512 against the trained PSN) & full, amplitude-only and phase-only memoriser of 1 or 5 training pilots on 20 or 106 unseen sequences & 1 \\
Coded & $2\times2$, TDL-A, 22 and 30~dB; 5G LDPC, rate 1/2 and 3/4; 256 codewords & plain, LIN-9, PA-aware LS, memoriser; pilots 42 and random & 1 \\
\bottomrule
\end{tabular*}
\end{table*}
\section{Conclusion}\label{sec:conclusion}

Pilots pass through the same amplifier as the data. If one pilot sequence
is used in every frame, the channel estimate carries the same small error
in every frame, and a learned receiver can learn this error instead of
learning to undo the amplifier. In simulated two- and four-antenna OFDM
links with a strongly driven amplifier and binary pilots, this made both
learned receivers we studied look better on their training pilot than on
any other pilot tested. At the highest signal-to-noise ratio tested, a network with
separate weights for each subcarrier lowered the bit error rate by 14 to
22\% on its training pilot and raised it by 2 to 18\% on every other pilot
tested, and longer training made the difference larger. A neural
receiver that follows an open-source tutorial gained 37 percentage
points more on its training pilot than on unseen pilots after short
training, and 9 points more after nine times longer training. This
remaining difference was present in every training run, but much of it
also appeared without amplifier distortion, so the amplifier is not its
only cause. In systems whose pilots change, such as 5G, the extra gain
would disappear.

Ordinary smoothing of the channel estimate across neighbouring
subcarriers, with no training and no knowledge of the amplifier, removed
most of the pilot error and lowered the bit error rate by about 30\% at
the highest signal-to-noise ratio tested. Of the learned receivers, only
the long-trained neural receiver did better on pilots it had not been
trained on, and only at high signal-to-noise ratio. When the channel spreads the signal over a long
delay, smoothing stops working and the problem returns.

Studies of learned receivers under amplifier distortion can avoid this
pitfall by training with changing pilots, testing on pilot sequences not
used in training, repeating the test with undistorted pilots, and
comparing with a smoothed channel estimate and other simple references.
Hardware tests with amplifier memory, NR reference signals in the full
link and other neural-receiver architectures remain to be done.



\section*{Data availability}
The simulation code (Sionna and PyTorch notebooks) will be provided upon request .

\section*{Declaration of generative AI and AI-assisted technologies in the manuscript preparation process}

During the preparation of this work, the author(s) used Claude Sonnet 5.5's assistance in writing the codes for the project,polishing the manuscript . After using this tool, the author(s) reviewed and edited the content as needed and take full responsibility for the content of the published article.


\end{document}